\documentclass[preprint,10pt]{elsarticle}

\usepackage{lineno,hyperref}
\usepackage{amsmath}
\usepackage{subfigure}
\usepackage{caption}
\usepackage{xcolor}
\usepackage{amssymb}
\modulolinenumbers[5]

\journal{Journal of COMPUT METHOD APPL M}

\begin{document}

\begin{frontmatter}

\title{A One-Field Monolithic Fictitious Domain\\
	 Method for Fluid-Structure Interactions} 


\author[]{Yongxing Wang\corref{mycorrespondingauthor}}
\cortext[mycorrespondingauthor]{Corresponding author}
\ead{scywa@leeds.ac.uk}

\author[]{Peter Jimack}
\author[]{Mark Walkley}

\address{School of Computing, University of Leeds, Leeds, UK, LS2 9JT}

\begin{abstract}
In this article, we present a one-field monolithic fictitious domain (FD) method for simulation of general fluid-structure interactions (FSI). ``One-field'' means only one velocity field is solved in the whole domain, based upon the use of an appropriate $L^2$ projection. ``Monolithic'' means the fluid and solid equations are solved synchronously (rather than sequentially). We argue that the proposed method has the same generality and robustness as FD methods with distributed Lagrange multiplier (DLM) but is significantly more computationally efficient (because of one-field) whilst being very straightforward to implement. The method is described in detail, followed by the presentation of multiple computational examples in order to validate it across a wide range of fluid and solid parameters and interactions.
\end{abstract}


\end{frontmatter}

\linenumbers

\section{Introduction }

Numerical simulation of fluid-structure interaction is a computational challenge because of its strong nonlinearity, especially when large deformation is considered. Based on how to couple the interaction between fluid and solid, existing numerical methods can be broadly categorized into two approaches: partitioned/segregated methods and monolithic/fully-coupled methods. Similarly, based on how to handle the mesh, they can also be broadly categorized into two further approaches: fitted mesh/conforming methods and unfitted/non-conforming mesh methods \cite{hou2012numerical}.

A fitted mesh means that the fluid and solid meshes match each other at the interface, and the nodes on the interface are shared by both the fluid and the solid, which leads to the fact that each interface node has both a fluid velocity and a solid velocity (or displacement) defined on it. It is apparent that the two velocities on each interface node should be consistent. There are typically two methods to handle this: partitioned/segregated methods \cite{kuttler2008fixed,Degroote_2009} and monolithic/fully-coupled methods \cite{Heil_2004,Heil_2008,Muddle_2012}. The former solve the fluid and solid equations sequentially and iterate until the velocities become consistent at the interface. These are more straightforward to implement but can lack robustness and may fail to converge when there is a significant energy exchange between the fluid and solid \cite{Degroote_2009}. The latter solve the fluid and solid equations simultaneously and often use a Lagrange Multiplier to weakly enforce the continuity of velocity on the interface \cite{Muddle_2012}. This has the advantage of achieving accurate and stable solutions, however the key computational challenge is to efficiently solve the large systems of nonlinear algebraic equations arising from the fully-coupled implicit discretization of the fluid and solid equations. Fitted mesh methods can accurately model wide classes of FSI problems, however maintaining the quality of the mesh for large solid deformations usually requires a combination of arbitrary Lagrangian-Eulerian (ALE) mesh movement and partial or full remeshing \cite{peterson1999solution}. These add to the computational expense and, when remeshing occurs, can lead to loss of conservation properties of the underlying discretization \cite{Walkley_2005}.

Unfitted mesh methods use two meshes to represent the fluid and solid separately and these do not generally conform to each other on the interface. In this case, the definition of the fluid problem may be extended to an augmented domain which includes the solid domain. Similarly to the fitted case, there are also two broad approaches to treat the solid domain: partitioned methods and monolithic methods. On an unfitted mesh, there is no clear boundary for the solid problem, so it is not easy to enforce the boundary condition and solve the solid equation. A wide variety of schemes have been proposed to address this issue, including the Immersed Finite Element Method (IFEM) \cite{zhang2004immersed,Zhang_2007,Wang_2009,Wang_2011,Wang_2013} and the Fictitious Domain method (FDM) \cite{Glowinski_2001,Yu_2005,baaijens2001fictitious,Kadapa_2016,Hesch_2014}. The IFEM developed from the Immersed Boundary method first introduced by Peskin \cite{peskin2002immersed}, and has had great success with applications in bioscience and biomedical fields. The classical IFEM does not solve solid equations at all. Instead, the solid equations are arranged on the right-hand side of the fluid equations as an FSI force, and these modified fluid equations are solved on the augmented domain (occupied by fluid and solid). There is also the Modified IFEM \cite{Wang_2013}, which solves the solid equations explicitly and iterates until convergence. The FDM has a similar spirit to IFEM in that it treats the domain occupied by solid as a fictitious/artificial fluid whose velocity/displacement is constrained to be the same as that of the solid. The FDM approach usually uses a distributed Lagrange multiplier (DLM) to enforce the constraint \cite{Glowinski_2001,Yu_2005,baaijens2001fictitious,Kadapa_2016} whilst the IFEM typically uses a pseudo body force which is evaluated from the known deformation of the solid and introduced into the fluid momentum equation. Reference \cite{Glowinski_2001} presents a fractional FD scheme for a rigid body interacting with the fluid, whilst \cite{Yu_2005} introduces a fractional step scheme using DLM/FD for fluid/flexible-body interactions. In the case of monolithic methods, \cite{baaijens2001fictitious} uses a FD/mortar approach to couple the fluid and structure, but the coupling is limited to a line (2D) representing the structure. Reference \cite{Hesch_2014} uses a mortar approach to solve fluid interactions with deformable and rigid bodies, and \cite{Kadapa_2016} also solves a fully-coupled FSI system with hierarchical B-Spline grids. There are also other monolithic methods based on unfitted meshes \cite{Robinson_Mosher_2011,Hachem_2013}. 

It can be seen that the major methods based on unfitted meshes either avoid solving the solid equations (IFEM) or solve them with additional variables (two velocity fields and Lagrange multiplier) in the solid domain. However, physically, there is only one velocity field in the solid domain. In this article, we follow the one-field spirit and only solve one velocity variable in the whole/augmented domain. We shall introduce a one-field FD method which can be categorized as a monolithic approach using an unfitted mesh.

In the one-field spirit, \cite{pironneau2016energy} introduces an Eulerian formulation by remeshing and \cite{Auricchio_2014} presents a 1D model using a one-field FD formulation but does not discuss how to compute the integrals arising from the two different domains. There are other similar Eulerian formulations for FSI problems, such as the eXtended Finite Element Method (XFEM) \cite{Gerstenberger_2008}, local modification of elements \cite{frei2016eulerian} and other fully Eulerian formulations \cite{Richter_2013,Wick_2013,Dunne} that are coupled with either local adaptivity or ALE methods. However these formulations are not in the spirit of one-field, usually the velocity of the fluid (including fictitious fluid), the displacement of the solid and the Lagrange multiplier are solved monolithically, which are three-field formulations (four fields if the moving mesh is solved for as well).

The main idea of the method presented here is as follows. (1) One-field formulation: we first discretize the control equations in time, re-write the solid equation in the form of a fluid equation (using the velocity as a variable rather than the displacement) and re-write the solid constitutive equation in the updated coordinate system. (2) $L^2$ projection (isoparametric interpolation): we then combine the fluid and solid equations and discretize them in an augmented domain. Finally the multi-physics problem is solved as a single field. 

The remaining sections are organized as follows. In section \ref{Governing equations}, the control equations and boundary conditions for fluid-structure interactions are introduced. Section \ref{Weak form} presents the weak form of the FSI system based on the augmented fluid domain. Section \ref{Discretization in time} introduces a splitting scheme after discretization in time. Section \ref{Linearization of the convection step} and \ref{Linearization of the diffusion step} discusses how to linearize the convection step and diffusion step respectively. In section \ref{Discretization in space}, the overall solution algorithm is presented after discretization in space, which clarifies one of the main differences of the proposed numerical scheme. In section \ref{Numerical experiments}, numerical examples are described to validate the proposed method across a wide range of flows and material. Some remarks and observations are discussed in section \ref{Discussion} and finally a brief summary is presented in section \ref{Conclusion and future works}.

\section{Governing equations for FSI}\label{Governing equations}
\begin{figure}[h!]
\centering
\includegraphics[width=3in,angle=0]{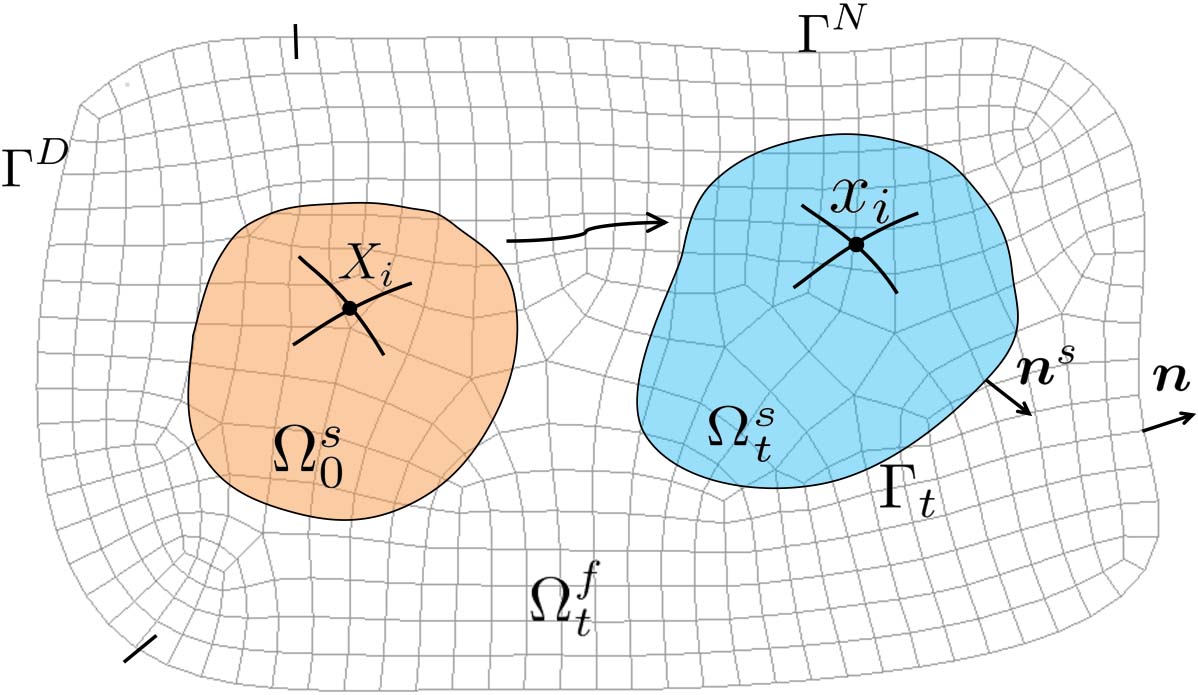}
\captionsetup{justification=centering}
\caption {\scriptsize Schematic diagram of FSI, $\Omega=\Omega_t^f\cup \Omega_t^s$, $\Gamma=\Gamma^D\cup \Gamma^N$.} 
\label{fig1}
\end{figure}

In the following context, let
\begin{equation}
\left(u,v\right)_\omega=\int_{\omega}uvd\omega,
\end{equation}
where $u$ and $v$ are functions defined in domain $\omega$. 
$\Omega_t^f\subset\mathbb{R}^d$ and $\Omega_t^s\subset\mathbb{R}^d$ (with $d=2$ in this article) denote the fluid and solid domain repectively which are time dependent regions as shown in Figure \ref{fig1}. $\Omega=\Omega_t^f \cup \Omega_t^s $ is a fixed domain and $\Gamma_t=\partial\Omega_t^f\cap\Omega_t^s$ is the moving interface between fluid and solid. All subscripts, such as $i$, $j$, and $k$, represent spatial dimension. If they are repeated in one term (including the bracket defined in (1)), it implies summation over the spatial dimension; if they are not repeated, they take the value from 1 to $d$. All superscripts are used to distinguish fluid and solid ($f$ and $s$ respectively), distinguish different boundaries ($\Gamma^D$ and $\Gamma^N$) or represent time step $\left(n\right)$. For example, $u_i^f$ and $u_i^s$ denote the velocity components of fluid and solid respectively, $\sigma_{ij}^f$ and $\sigma_{ij}^s$ denote the stress tensor components of fluid and solid respectively, and $\left(u_i^s\right)^n$ is a solid velocity component at time $t^n$. Matrices and vectors are denoted by bold letters.

In our model we assume an incompressible fluid governed by the following equations in $\Omega_t^f$ as shown in Figure \ref{fig1}:
\begin{equation} \label{fluid1}
\rho^f\frac{Du_i^f}{Dt}
-\frac{\partial \sigma_{ij}^f}{\partial x_j}=\rho^f g_i,
\end{equation}

\begin{equation} \label{fluid2}
\frac{\partial u_j^f}{\partial x_j}=0,
\end{equation}

\begin{equation} \label{fluid3}
\sigma_{ij}^f=\mu^f \left(\frac{\partial u_i^f}{\partial x_j}+\frac{\partial u_j^f}{\partial x_i} \right)-p^f\delta_{ij}
=\tau_{ij}^f-p^f\delta_{ij}.
\end{equation}

We also assume an incompressible solid that is governed by the following equations in $\Omega_t^s$ as shown in Figure \ref{fig1}:
\begin{equation} \label{solid1}
\rho^s \frac{Du_i^s}{Dt} -\frac{\partial \sigma_{ij}^s}{\partial x_j}=\rho^s g_i,
\end{equation}

\begin{equation} \label{solid2}
\frac{\partial u_j^s}{\partial x_j}=0,
\end{equation}

\begin{equation} \label{solid3}
\sigma_{ij}^s=\mu^s \left(\frac{\partial x_i^s}{\partial X_k}\frac{\partial x_j^s}{\partial X_k} - \delta_{ij} \right)-p^s\delta_{ij}
=\tau_{ij}^s-p^s\delta_{ij}.
\end{equation}

In the above $\tau_{ij}^f$ and $\tau_{ij}^s$ are the deviatoric stress of the fluid and solid respectively, $\rho^f$ and $\rho^s$ are the density of the fluid and solid respectively, $\mu^f$ is the fluid viscosity, and $g_i$ is the acceleration due to gravity. Note that (\ref{solid1})-(\ref{solid3}) describe an incompressible neo-Hookean model that is based on \cite{baaijens2001fictitious} and is suitable for large displacements. In this model, $\mu^s$ is the shear modulus and $p^s$ is the pressure of the solid ($p^f$ being the fluid pressure in (\ref{fluid3})). We denote by $x_i$ the current coordinates of the solid or fluid, and by $X_i$ the reference coordinates of the solid, whilst ${\bf F}=\left[\frac{\partial x_i}{\partial X_j}\right]$ is the deformation tensor of the solid and $\frac{D}{Dt}$ represents the total derivative of time.

On the interface boundary $\Gamma_t$: 
 \begin{equation}\label{interfaceBC1}
 u_i^f=u_i^s,
 \end{equation}
 \begin{equation}\label{interfaceBC2}
 \sigma_{ij}^fn_j^s=\sigma_{ij}^sn_j^s,
 \end{equation}
where $n_j^s$ denotes the component of outward pointing unit normal, see Figure \ref*{fig1}.

Dirichlet and Neumann boundary conditions may be imposed for the fluid:

\begin{equation}
u_i^f=\bar{u}_i \quad on \quad  \Gamma^D,
\end{equation}

\begin{equation}\label{NeummanBC}
\sigma_{ij}^fn_j=\bar{h}_i \quad on \quad  \Gamma^N.
\end{equation}

Finally, initial conditions are typically set as:
\begin{equation} \label{initialcd}
 \left. u_i^f\right|_{t=0}=\left. u_i^s\right|_{t=0}=0,
\end{equation}
though they may differ from (\ref {initialcd}).

\section{Weak formulation}\label{Weak form}
Let
$
u_i=\left \{ 
\begin{matrix}
{u_i^f \quad in \quad \Omega_t^f} \\
{u_i^s \quad in \quad \Omega_t^s} \\
\end{matrix}\right.
$
and
$
p=\left \{ 
\begin{matrix}
{p^f \quad in \quad \Omega_t^f} \\
{p^s \quad in \quad \Omega_t^s} \\
\end{matrix}\right.
$
.
We then perform the following symbolic operations:
$$
\left({\rm Eq.}(\ref{fluid1}),v_i\right)_{\Omega_t^f}-\left({\rm Eq.}(\ref{fluid2}),q\right)_{\Omega_t^f}
+\left({\rm Eq.}(\ref{solid1}),v_i\right)_{\Omega_t^s}-\left({\rm Eq.}(\ref{solid2}),q\right)_{\Omega_t^s},
$$
for independent test functions $v_i\in H_0^1(\Omega)$ and $q\in L^2(\Omega)$.

Integrating the stress terms by parts, using constitutive equations (\ref{fluid3}) and (\ref{solid3}) and boundary condition (\ref{NeummanBC}), gives the following weak form for the FSI system.

Find $u_i\in H^1(\Omega)$ and $p\in L_0^2\left(\Omega\right)$ such that
\begin{equation}\label{weakaugmented}
\begin{split}
& \rho^f\left( \frac{D{u_i}}{Dt}, v_i \right)_\Omega
+\left(\tau_{ij}^f, \frac{\partial v_i}{\partial x_j}\right)_\Omega
-\left(p,\frac{\partial v_j}{\partial x_j}\right)_\Omega
-\left(\frac{\partial u_j}{\partial x_j}, q\right)_\Omega  \\
& +\left(\rho^s-\rho^f\right)\left( \frac{D{u_i}}{Dt}, v_i\right)_{\Omega_t^s}
+\left(\tau_{ij}^s-\tau_{ij}^f, \frac{\partial v_i}{\partial x_j}\right)_{\Omega_t^s} \\
&= \left(\bar{h}_i, v_i\right)_{\Gamma^N}
+\rho^f\left(g_i, v_i\right)_{\Omega}
+\left(\rho^s-\rho^f\right)\left(g_i, v_i\right)_{\Omega_t^s},
\end{split}
\end{equation}
$\forall v_i \in H_0^1(\Omega)$ and $\forall q \in L^2\left({\Omega}\right)$. In the above, $\rho^f$ and $\tau_{ij}^f$ are extended to be defined over the whole of $\Omega$, and $L_0^2(\Omega)=\left\{p:p\in L^2(\Omega),\left.p\right|_{P_0}=0\right\}$, where $P_0$ is a reference point.
Note that the integrals on the interface (boundary forces) are cancelled out using boundary condition (\ref{interfaceBC2}). This is not surprising because they are internal forces for the whole FSI system considered here.

{\bf Remark 1} The fluid deviatoric stress $\tau_{ij}^f$ is generally far smaller than the solid deviatoric stress $\tau_{ij}^s$, so we choose to neglect the fluid deviatoric stress $\tau_{ij}^f$ in $\Omega^s$ in what follows. Note that the classical IFEM neglects the whole fluid stress $\sigma_{ij}^f$ when computing the FSI force \cite{zhang2004immersed}. An equivalent way of interpreting neglecting $\tau_{ij}^f$ in $\Omega^s$ is to view the solid as being slightly visco-elastic, having the same viscosity as the fluid.

\section{Discretization in time}\label{Discretization in time}
The integrals in equation (\ref{weakaugmented}) are carried out in two different domains as illustrated in Figure \ref{fig1}. We use an Eulerian mesh to represent $\Omega$ and an updated Lagrangian mesh to represent $\Omega^s$, therefore the total time derivatives in these two different domains have different expressions, i.e: 
\begin{equation}\label{derivativef}
\frac{Du_i}{Dt}=\frac{\partial u_i}{\partial t}+u_j\frac{\partial u_i}{\partial x_j} \quad in \quad \Omega,
\end{equation}
and
\begin{equation}\label{derivatives}
\frac{Du_i^s}{Dt}=\frac{\partial u_i^s}{\partial t} \quad in \quad \Omega^s.
\end{equation}

Firstly, based on the above two equations (\ref{derivativef}) and (\ref{derivatives}), we discretize (\ref{weakaugmented}) in time using a backward finite difference. Then omitting the superscript $n+1$, showing the solution is at the end of the time step, for convenience, we obtain:
\begin{equation}\label{weak2}
\begin{split}
& \rho^f\left( \frac{u_i-u_i^n}{\Delta t}+u_j\frac{\partial u_i}{\partial x_j}, v_i \right)_\Omega
+\left(\tau_{ij}^f, \frac{\partial v_i}{\partial x_j}\right)_\Omega
-\left(p,\frac{\partial v_j}{\partial x_j}\right)_\Omega-\left(\frac{\partial u_j}{\partial x_j}, q\right)_\Omega  \\
& +\left(\rho^s-\rho^f\right)\left( \frac{u_i-u_i^n}{\Delta t}, v_i\right)_{\Omega_{n+1}^s}
+\left(\tau_{ij}^s, \frac{\partial v_i}{\partial x_j}\right)_{\Omega_{n+1}^s} \\
&= \left(\bar{h}_i, v_i\right)_{\Gamma^N}
+\rho^f\left(g_i, v_i\right)_{\Omega}
+\left(\rho^s-\rho^f\right)\left(g_i, v_i\right)_{\Omega_{n+1}^s}.
\end{split}
\end{equation}
Note that in the above we have replaced $\Omega_{t^{n+1}}^s$ by $\Omega_{n+1}^s$ for notational convenience. Using the splitting method of \cite[Chapter 3]{Zienkiewic2014}, equation (\ref{weak2}) can be expressed in the following two steps.

(1) Convection step:
\begin{equation}\label{weakconvection}
\rho^f\left(\frac{u_i^*-u_i^n}{\Delta t}+u_j^*\frac{\partial u_i^*}{\partial x_j}, v_i\right)_\Omega=0;
\end{equation}

(2) Diffusion step:
\begin{equation}\label{weakdiffusion}
\begin{split}
& \rho^f\left( \frac{u_i-u_i^*}{\Delta t}, v_i \right)_\Omega
+\left(\tau_{ij}^f, \frac{\partial v_i}{\partial x_j}\right)_\Omega
-\left(p,\frac{\partial v_j}{\partial x_j}\right)_\Omega-\left(\frac{\partial u_j}{\partial x_j}, q\right)_\Omega  \\
& +\left(\rho^s-\rho^f\right)\left( \frac{u_i-u_i^n}{\Delta t}, v_i \right)_{\Omega_{n+1}^s}
+\left(\tau_{ij}^s, \frac{\partial v_i}{\partial x_j}\right)_{\Omega_{n+1}^s} \\
&= \left(\bar{h}_i, v_i\right)_{\Gamma^N}
+\rho^f\left(g_i, v_i\right)_{\Omega}
+\left(\rho^s-\rho^f\right)\left(g_i, v_i\right)_{\Omega_{n+1}^s}.
\end{split}
\end{equation}
The treatment of the above two steps is described separately in the following subsections.

\section{Linearization of the convection step}\label{Linearization of the convection step}
In this section, two methods are introduced to treat the convection equation: Least-squares method and Taylor-Galerkin method, both of which can be used in the framework of the proposed scheme. Some numerical results for comparison between these two methods are discussed subsequently in section 5. Because the overall scheme is explicit, all non-linear terms are linearized using the values from the last time step. Of course, the scheme can be made implicit with the same linearized form by iterating within each time step starting from the value at the last time step.

\subsection{Least-squares method}\label{Least-squares method}
Linearization of equation (\ref{weakconvection}) gives,
\begin{equation}
\left(u_i^*+\Delta t\left(u_j^*\frac{\partial u_i^n}{\partial x_j}+u_j^n\frac{\partial u_i^*}{\partial x_j}\right), v_i\right)_\Omega
=\left(u_i^n+\Delta t u_j^n\frac{\partial u_i^n}{\partial x_j}, v_i\right)_\Omega.
\end{equation}
For Least-squares method \cite{bochev2009least}, we may choose the test function in the following form:
\begin{equation}
v_i=L\left(w_i\right)=w_i+\Delta t\left(w_j\frac{\partial u_i^n}{\partial x_j}+u_j^n\frac{\partial w_i}{\partial x_j}\right)_{\Omega},
\end{equation}
where $w_i \in H_0^1(\Omega)$. In such a case, the weak form of (\ref{weakconvection}) is:
\begin{equation}\label{lastweakofconvection}
\left(L\left(u_i^*\right), L\left(w_i\right)\right)_\Omega
=\left(u_i^n+\Delta t u_j^n\frac{\partial u_i^n}{\partial x_j}, L\left(w_i\right)\right)_\Omega.
\end{equation}
In our method a standard biquadratic finite element space is used to discretize equation (\ref{lastweakofconvection}) directly.

\subsection{Taylor-Galerkin method}
It is also possible to linearize equation (\ref{weakconvection}) as:
\begin{equation}\label{cbsconvection1}
\left(\frac{u_i^*-u_i^n}{\Delta t}+\frac{1}{2}u_j^n\frac{\partial}{\partial x_j}\left(u_i^*+u_i^n\right), v_i\right)_\Omega=0
\end{equation}
or
\begin{equation}\label{cbsconvection2}
\left(\frac{u_i^*-u_i^n}{\Delta t}+u_j^n\frac{\partial u_i^n}{\partial x_j}, v_i\right)_\Omega=0.
\end{equation}
Rewriting (\ref{cbsconvection2}) as
\begin{equation}\label{cbsconvection-strong-form}
u_i^*=u_i^n-{\Delta t}u_j^n\frac{\partial u_i^n}{\partial x_j},
\end{equation}
substituting (\ref{cbsconvection-strong-form}) into equation (\ref{cbsconvection1}), and applying integration by parts we obtain:
\begin{equation}\label{cbs-last2}
\left(\frac{u_i^*-u_i^n}{\Delta t}+u_j^n\frac{\partial u_i^n}{\partial x_j}, v_i\right)_\Omega
=-\frac{\Delta t}{2}\left(u_k^n\frac{\partial u_i^n}{\partial x_k}, u_j^n\frac{\partial v_i}{\partial x_j}\right)_\Omega,
\end{equation}
where the boundary integral is neglected because $u_i^n$ is the solution of the previous diffusion step, which means no convection exists on the boundary after the diffusion step. Finally the weak form of Taylor-Galerkin method \cite[Chapter 2]{Zienkiewic2014} can be expressed, by rearranging the last equation, as:
\begin{equation}\label{cbs-last}
\left(u_i^*,v_i\right)_\Omega=
\left(u_i^n-\Delta t u_j^n\frac{\partial u_i^n}{\partial x_j},v_i\right)_\Omega
-\frac{\Delta t^2}{2}\left(u_k^n\frac{\partial u_i^n}{\partial x_k}, u_j^n\frac{\partial v_i}{\partial x_j}\right)_\Omega.
\end{equation}

\section{Linearization of the diffusion step}\label{Linearization of the diffusion step}
As mentioned above, the overall scheme is explicit, so all the derivatives are computed on the known coordinate $\left(x_i^s\right)^n$ (denoted $x_i^n$ for convenience). One could also construct $x_i^{n+1}$ at each time step and take derivatives with respect to $x_i^{n+1}$, however we do not consider such an approach in this article. According to the definition of $\tau_{ij}^s$ in equation (\ref{solid3}),
\begin{equation}
\left(\tau_{ij}^s\right)^{n+1}=\mu^s\left(\frac{\partial x_i^{n+1}}{\partial X_k}
\frac{\partial x_j^{n+1}}{\partial X_k}-\delta_{ij}\right).
\end{equation}
The last equation, using a chain rule, can also be expressed as:
\begin{equation}
\begin{split}
&\left(\tau_{ij}^s\right)^{n+1}=\mu^s\left(\frac{\partial x_i^{n+1}}{\partial x_k^n}
\frac{\partial x_j^{n+1}}{\partial x_k^n}-\delta_{ij}\right) \\
&+\mu^s\frac{\partial x_i^{n+1}}{\partial x_k^n}\left(\frac{\partial x_k^n}{\partial X_m}
\frac{\partial x_l^n}{\partial X_m}-{\delta_{kl}}
\right)\frac{\partial x_j^{n+1}}{\partial x_l^n}
\end{split}
,
\end{equation}
and then $\left(\tau_{ij}^s\right)^{n+1}$ can be expressed by coordinate $x_i^n$ as follows:
\begin{equation}
\begin{split}
&\left(\tau_{ij}^s\right)^{n+1}=\mu^s\left(\frac{\partial x_i^{n+1}}{\partial x_k^n}
\frac{\partial x_j^{n+1}}{\partial x_k^n}-\delta_{ij}\right)
+\frac{\partial x_i^{n+1}}{\partial x_k^n}\left(\tau_{kl}^s\right)^n
\frac{\partial x_j^{n+1}}{\partial x_l^n}
\end{split}
.
\end{equation}
Using $x_i^{n+1}-x_i^n=u_i^{n+1}\Delta t$, which is the displacement at the current step, the last equation may be expressed as:
\begin{equation}\label{solidstress}
\begin{split}
&\left(\tau_{ij}^s\right)^{n+1}=\mu^s \Delta t\left(\frac{\partial u_i^{n+1}}{\partial x_j^n}+\frac{\partial u_j^{n+1}}{\partial x_i^n}+\Delta t\frac{\partial u_i^{n+1}}{\partial x_k^n}\frac{\partial u_j^{n+1}}{\partial x_k^n}\right)+\left(\tau_{ij}^s\right)^n   \\
&+\Delta t^2\frac{\partial u_i^{n+1}}{\partial x_k^n}\left(\tau_{kl}^s\right)^n\frac{\partial u_j^{n+1}}{\partial x_l^n}
+\Delta t\frac{\partial u_i^{n+1}}{\partial x_k^n}\left(\tau_{kj}^s\right)^n
+\Delta t\left(\tau_{il}^s\right)^n\frac{\partial u_j^{n+1}}{\partial x_l^n}.
\end{split}
\end{equation}
Finally, after linearization of the last equation, the weak form (\ref{weakdiffusion}) can be expressed as:
\begin{equation}\label{lastbigequation}
\begin{split}
&\rho^f\left(\frac{u_i-u_i^*}{\Delta t}, v_i\right)_\Omega+\left(\rho^s-\rho^f\right)\left(\frac{u_i^s-\left(u_i^s\right)^n}{\Delta t}, v_i\right)_{\Omega_{n+1}^s} \\
&+\mu^f\left(\frac{\partial u_i}{\partial x_j}+\frac{\partial u_j}{\partial x_i}, \frac{\partial v_i}{\partial x_j}\right)_\Omega
-\left(p, \frac{\partial v_j}{\partial x_j}\right)_\Omega-\left(\frac{\partial u_j}{\partial x_j}, q\right)_\Omega \\
&+\mu^s\Delta t\left(\frac{\partial u_i}{\partial x_j}+\frac{\partial u_j}{\partial x_i}
+\Delta t\frac{\partial u_i}{\partial x_k}\frac{\partial u_j^n}{\partial x_k}
+\Delta t\frac{\partial u_i^n}{\partial x_k}\frac{\partial u_j}{\partial x_k}, 
\frac{\partial v_i}{\partial x_j}  \right)_{\Omega_{n+1}^s} \\
&+\Delta t^2\left(\frac{\partial u_i}{\partial x_k}\left(\tau_{kl}^s\right)^n\frac{\partial u_j^n}{\partial x_l}  
+\frac{\partial u_i^n}{\partial x_k}\left(\tau_{kl}^s\right)^n\frac{\partial u_j}{\partial x_l}, 
\frac{\partial v_i}{\partial x_j}\right)_{\Omega_{n+1}^s} \\
&+\Delta t\left(\frac{\partial u_i}{\partial x_k}\left(\tau_{kj}^s\right)^n+\left(\tau_{il}^s\right)^n\frac{\partial u_j}{\partial x_l}, \frac{\partial v_i}{\partial x_j}\right)_{\Omega_{n+1}^s} \\
&=\left(\bar{h}_i, v_i\right)_{\Gamma^N}+\rho^f\left(g_i, v_i\right)_\Omega+\left(\rho^s-\rho^f\right)\left(g_i, v_i\right)_{\Omega_{n+1}^s} \\
&+\left(\mu^s\Delta t^2\frac{\partial u_i^n}{\partial x_k}\frac{\partial u_j^n}{\partial x_k}
+\Delta t^2\frac{\partial u_i^n}{\partial x_k}\left(\tau_{kl}^s\right)^n\frac{\partial u_j^n}{\partial x_l}-\left(\tau_{ij}^s\right)^n, \frac{\partial v_i}{\partial x_j}
\right)_{\Omega_{n+1}^s}.
\end{split}
\end{equation}

The spatial discretization of the above linearized weak form will be discussed in the following section, along with the overall solution algorithm.

\section{Discretization in space and solution algorithm} \label{Discretization in space}
\subsection{Spatial discretization}
We shall use a fixed Eulerian mesh for $\Omega$ and an updated Lagrangian mesh for $\Omega_{n+1}^s$ to discretize equation (\ref{lastbigequation}). First, we discretize $\Omega$ as $\Omega^h$ using ${\bf P}_2{\rm P}_1$ elements (the Taylor-Hood element) with the corresponding finite element spaces as
$$
V^h(\Omega^h)=span\left\{\varphi_1,\cdots,\varphi_{N^u}\right\} \subset H^1\left(\Omega\right)
$$
and
$$
L^h(\Omega^h)=span\left\{\phi_1,\cdots,\phi_{N^p}\right\} \subset L^2\left(\Omega\right).
$$
The approximated solution ${\bf u}^h$ and $p^h$ can be expressed in terms of these basis functions as
\begin{equation}\label{uh}
{\bf u}^h({\bf x})=\sum_{i=1}^{N^u}{\bf u}({\bf x}_i)\varphi_i({\bf x}), \quad
p^h({\bf x})=\sum_{i=1}^{N^p}p({\bf x}_i)\phi_i({\bf x}).
\end{equation}

We further discretize $\Omega_{n+1}^s$ as $\Omega_{n+1}^{sh}$ (actually it is discretized once on $\Omega_0^s$ and then updated from the previous mesh) using ${\rm P}_1$ elements (bilinear triangle element) with the corresponding finite element spaces as:
$$
V^{sh}(\Omega_{n+1}^{sh})=span\left\{\varphi_1^s,\cdots,\varphi_{N^s}^s\right\} \subset H^1\left(\Omega_{n+1}^s\right),
$$
and approximate $\left.{\bf u}^h({\bf x})\right|_{{\bf x}\in\Omega_{n+1}^{sh}}$ as:
\begin{equation}\label{ush}
{\bf u}^{sh}\left({\bf x}\right)
=\sum_{i=1}^{N^s}{\bf u}^h({\bf x}_i^s)\varphi_i^s({\bf x})
=\sum_{i=1}^{N^s}\sum_{j=1}^{N^u}{\bf u}({\bf x}_j)\varphi_j({\bf x}_i^s)\varphi_i^s({\bf x}),
\end{equation}
where ${\bf x}_i^s$ is the nodal coordinate of the solid mesh. Notice that the above approximation defines an $L^2$ projection $P_{n+1}$ from $V^h$ to $V^{sh}$: $P_{n+1}\left({\bf u}^h({\bf x})\right)={\bf u}^{sh}\left({\bf x}\right)$.

Substituting (\ref{uh}), (\ref{ush}) and similar expressions for the test functions ${\bf v}^h$, $q^h$ and ${\bf v}^{sh}$ into equation (\ref{lastbigequation}) gives the following matrix form:
\begin{equation}\label{lastlinerequation}
\begin{bmatrix}
{\bf A} & {\bf B} \\
{\bf B}^{\rm T} &{\bf 0}
\end{bmatrix}
\begin{pmatrix}
{\bf u} \\
{\bf p}
\end{pmatrix}
=
\begin{pmatrix}
{\bf b} \\
{\bf 0}
\end{pmatrix}
,
\end{equation}
where
\begin{equation}\label{matirx_A}
{\bf A}={\bf M}/\Delta t+{\bf K}+{\bf D}^{\rm T}\left({\bf M}^s/\Delta t+{\bf K}^s\right){\bf D},
\end{equation}
and
\begin{equation}\label{forcevector_b}
{\bf b}={\bf f}+{\bf D}^{\rm T}{\bf f}^s+{\bf M}{\bf u}^*/\Delta t+{\bf D}^{\rm T}{\bf M}^s{\bf D}{\bf u}^n/\Delta t.
\end{equation}

In the above, matrix ${\bf D}$ is the isoparametric interpolation matrix derived from equation (\ref{ush}) which can be expressed as
\begin{equation*}
{\bf D}=
\begin{bmatrix}
{\bf P}^{\rm T} & {\bf 0} \\
{\bf 0} & {\bf P}^{\rm T} \\
\end{bmatrix}
,
{\bf P}_{ij}=\varphi_i({\bf x}_j^s)
.
\end{equation*}
All the other matrices and vectors arise from standard FEM discretization: ${\bf M}$ and ${\bf M}^s$ are mass matrices from discretization of integrals in $\Omega^h$ (with shape function $\varphi_i$) and $\Omega^{sh}$ (with shape function $\varphi_i^s$) respectively, and similarly for stiffness matrices ${\bf K}$ and ${\bf K}^s$. ${\bf B}$ is from discretization of integral $-\left(p,\frac{\partial v_j}{\partial v_j}\right)$ in (\ref{lastbigequation}). The force vectors ${\bf f}$ and ${\bf f}^s$ come from discretization of integrals on the right-hand side of (\ref{lastbigequation}) in $\Omega^h$ and $\Omega^{sh}$ respectively. The specific expressions of these matrices and vectors can be found in \ref{expressions_of_maatrices_and_vectors}.

\subsection{Overall solution algorithm}
Having derived a discrete system of equations we now describe the solution algorithm at each time step.
\begin{enumerate}[(1)]
	\item Given the solid configuration $\left({\bf x}^s\right)^n$ and velocity field ${\bf u}^n=\left\{\begin{matrix}
	\left({\bf u}^f\right)^n & in \quad \Omega^f \\
	\left({\bf u}^s\right)^n & in \quad \Omega^s
	\end{matrix}\right.$ at time step $n$.
	\item Discretize the convection equation (\ref{lastweakofconvection}) or (\ref{cbs-last}) and solve it to get an intermediate velocity ${\bf u}^*$.
	\item Compute the interpolation matrix  and solve equation (\ref{lastlinerequation}) using ${\bf u}^*$ and $\left({\bf u}^s\right)^n$ as initial values to get velocity field ${\bf u}^{n+1}$.
    \item Compute solid velocity $\left({\bf u}^s\right)^{n+1}={\bf D}{\bf u}^{n+1}$ and update the solid mesh by $\left({\bf x}^s\right)^{n+1}=\left({\bf x}^s\right)^n+\Delta t\left({\bf u}^s\right)^{n+1}$, then go to step (1) for the next time step.
\end{enumerate}

{\bf Remark 2} The choice of ${\rm P}_1$ element for an updated domain $\Omega^s$ is convenient, because the form of the bilinear shape functions stays the same when updating the nodal coordinates using $\left({\bf x}^s\right)^{n+1}=\left({\bf x}^s\right)^n+\Delta t\left({\bf u}^s\right)^{n+1}$.

{\bf Remark 3} When implementing the algorithm, it is unnecessary to perform the matrix multiplication ${\bf D}^{\rm T}{\bf K}^s{\bf D}$ globally, because the FEM interpolation is locally based. All the matrix operations can be computed based on the local element matrices only. Alternatively, if an iterative solver is used, it is actually unnecessary to compute ${\bf D}^{\rm T}{\bf K}^s{\bf D}$. What an iterative step needs is to compute $\left({\bf D}^{\rm T}{\bf K}^s{\bf D}\right){\bf u}$ for a given vector ${\bf u}$, therefore one can compute ${\bf Du}$ first, then ${\bf K}^s\left({\bf D}{\bf u}\right)$, and last ${\bf D}^{\rm T}\left({\bf K}^s{\bf D}{\bf u}\right)$.

\section{Numerical experiments}\label{Numerical experiments}

In this section, we present some numerical examples that have been selected to allow us to assess the accuracy and the versatility of our proposed numerical scheme. We demonstrate convergence in time and space, furthermore, we favorably compare results with those obtained using monolithic approaches and IFEM, as well as compare against results from laboratory experiment. 

In order to improve the computational efficiency, an adaptive spatial mesh with hanging nodes is used in all the following numerical experiments. Readers can reference \cite{Gupta_1978,Fries_2010,Bangerth_2009,Zander_2015} for details of the treatment of hanging nodes. The Least-squares method (section \ref{Least-squares method}) is used to treat the convection step in all tests unless stated otherwise.

\subsection{Oscillation of a flexible leaflet oriented across the flow direction}\label{Oscillation of a flexible leaflet oriented across the flow direction}
\begin{figure}[h!]
	\centering
	\includegraphics[width=4in,angle=0]{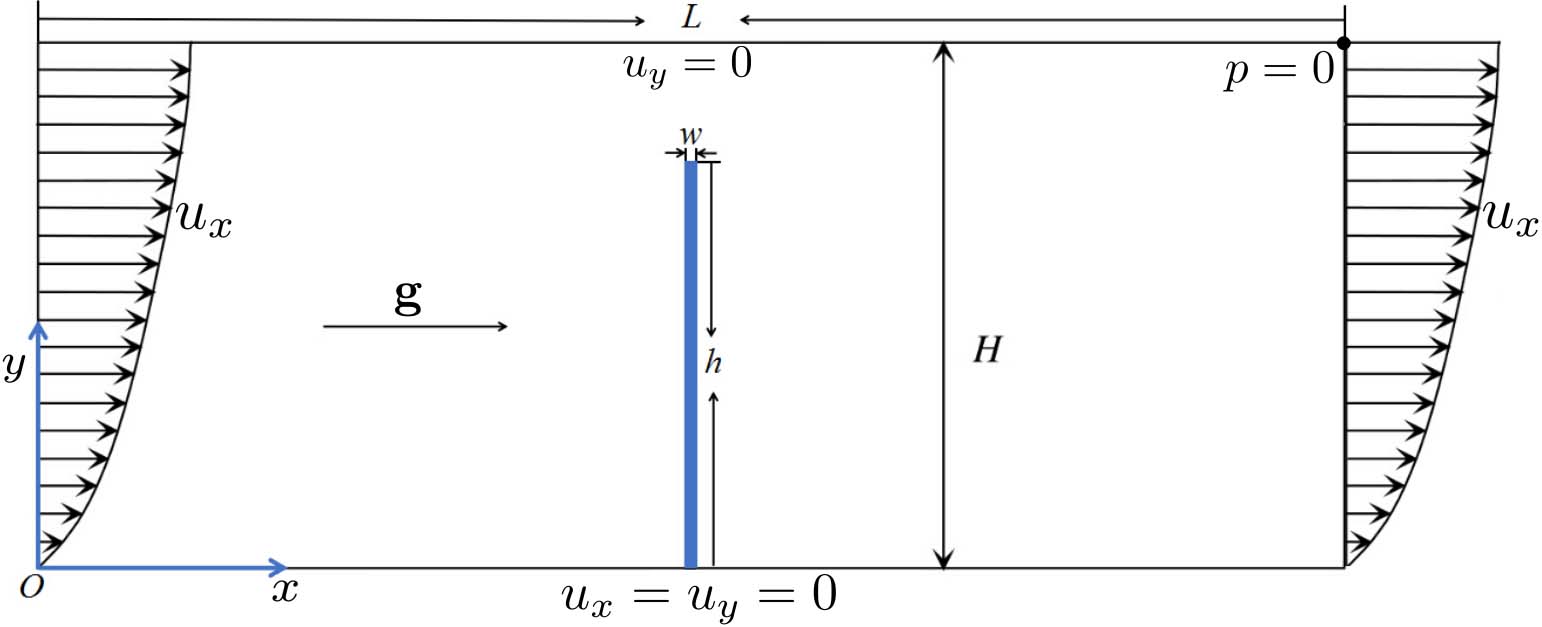}
	\captionsetup{justification=centering}
	\caption {\scriptsize Computational domain and boundary conditions, taken from \cite{baaijens2001fictitious}.} 
	\label{fig2}
\end{figure}

This numerical example is used by \cite{Yu_2005,baaijens2001fictitious,Kadapa_2016} to validate their methods. We first use the same parameters as used in the above three publications in order to compare results and test convergence in time and space. We then use a wide range of parameters to show the robustness of our method. The computational domain and boundary conditions are illustrated in Figure \ref{fig2}.

The inlet flow is in the x-direction and given by $u_x=15.0y\left(2-y\right)sin\left(2\pi t\right)$. Gravity is not considered in the first test (i.e. ${\bf g}=0$), and other fluid and solid properties are presented in Table \ref{Properties and domain size for test problem with a leaflet in a channel}.
\begin{table}[h!]
	\centering
	\begin{tabular}{|c|c|}
		\hline
		Fluid  & Leaflet \\
		\hline
		$L=4.0$ $m$ & $w=0.0212$ $m$ \\
		$H=1.0$ $m$ & $h=0.8$ $m$ \\
		$\rho^f=100$ $\left.kg\right/{m^3}$ & $\rho^s=100$ $\left.kg\right/{m^3}$ \\
		$\mu^f=10$ $\left.N\cdot s\right/{m^2}$ & $\mu^s=10^7$ $\left.N\right/{m^2}$ \\
		\hline
	\end{tabular}
	\captionsetup{justification=centering}
	\caption{Properties and domain size for test problem \ref{Oscillation of a flexible leaflet oriented across the flow direction}\\ with a leaflet oriented across the flow direction.}
	\label{Properties and domain size for test problem with a leaflet in a channel}
\end{table}

\begin{figure}[h!]
\centering
    \subfigure[h][$t=0.1s$]{ 
	\includegraphics[width=4.7in,angle=0]{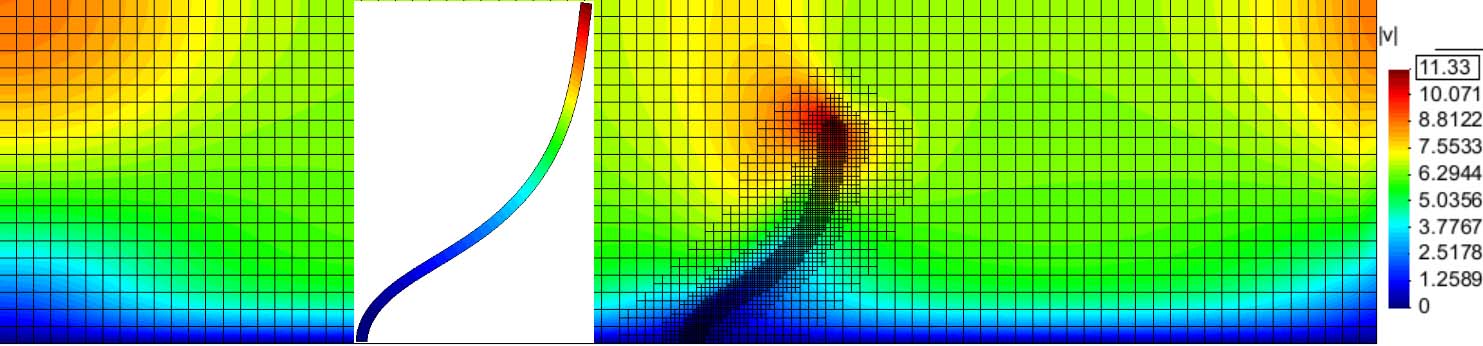}
    }
    \subfigure[$t=0.2s$]{ 
    \includegraphics[width=4.7in,angle=0]{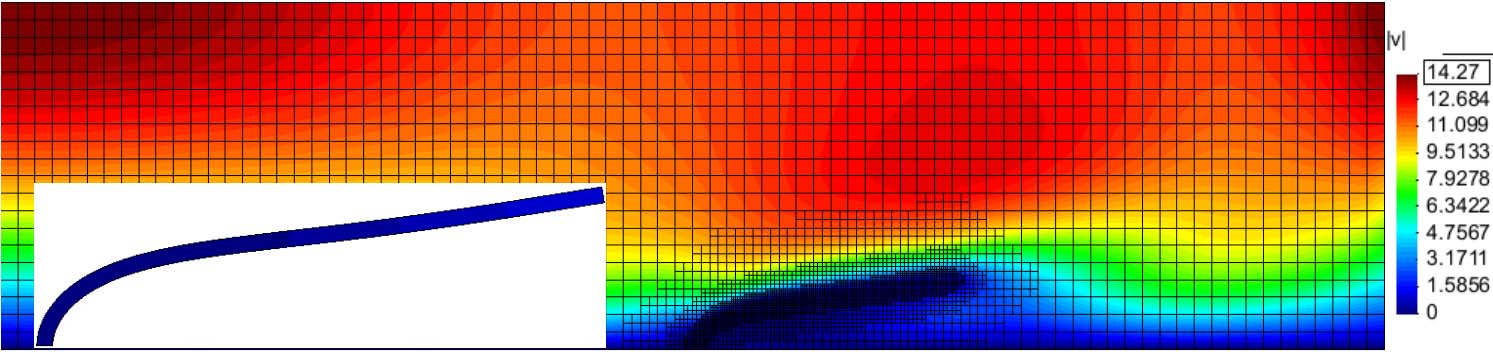}
    }    
    \subfigure[$t=0.6s$]{ 
    	\includegraphics[width=4.7in,angle=0]{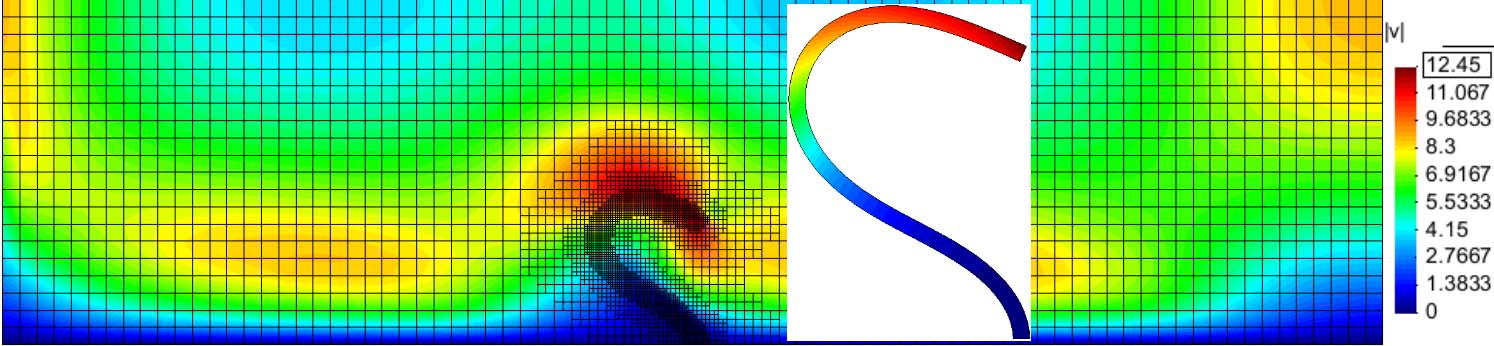}
    }        
    \subfigure[$t=0.8s$]{ 
    	\includegraphics[width=4.7in,angle=0]{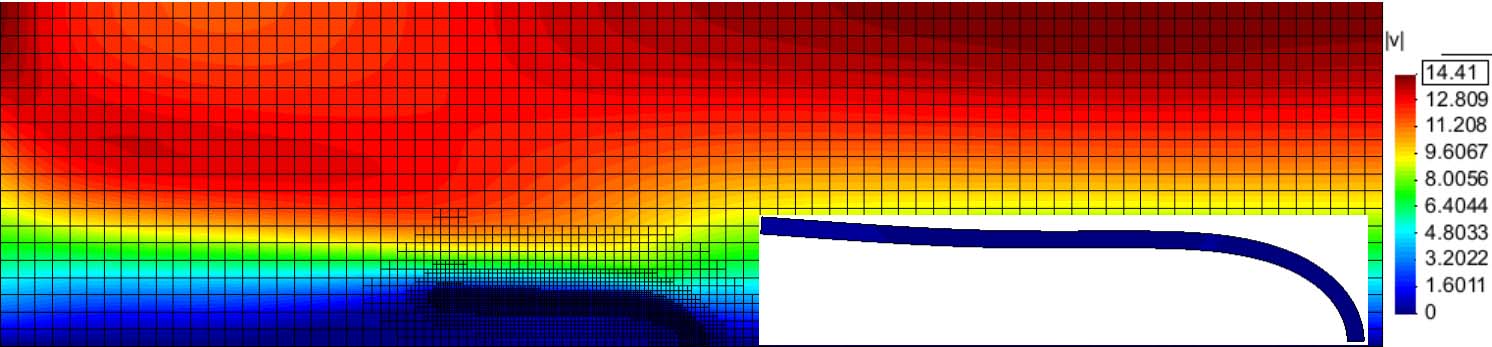}
    }     
\captionsetup{justification=centering}
\caption {\scriptsize Configuration of leaflet and magnitude of velocity on the adaptive fluid mesh.} 
\label{Configuration of leaflet and magnitude of velocity on the adaptive fluid mesh}
\end{figure}

The leaflet is approximated with 1200 linear triangles with 794 nodes (medium mesh size), and the corresponding fluid mesh is adaptive in the vicinity of the leaflet so that it has a similar size. A stable time step $\Delta t=5.0\times 10^{-4}s$ is used in these initial simulations. The configuration of the leaflet is illustrated at different times in Figure \ref{Configuration of leaflet and magnitude of velocity on the adaptive fluid mesh}.

\begin{figure}[h!]
	\begin{minipage}[t]{0.5\linewidth}
    \centering  
    \includegraphics[width=2.2in,angle=0]{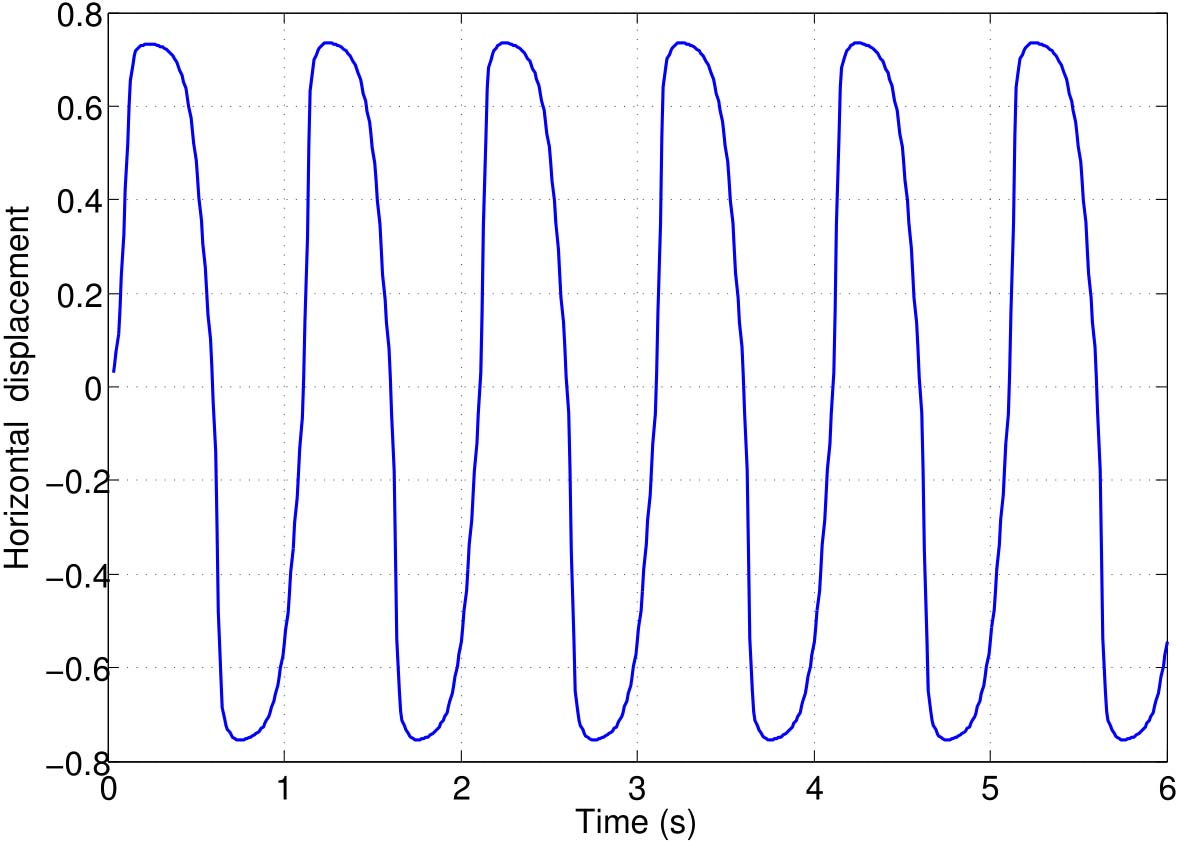}
    \end{minipage}
	\begin{minipage}[t]{0.5\linewidth}
		\centering  
		\includegraphics[width=2.2in,angle=0]{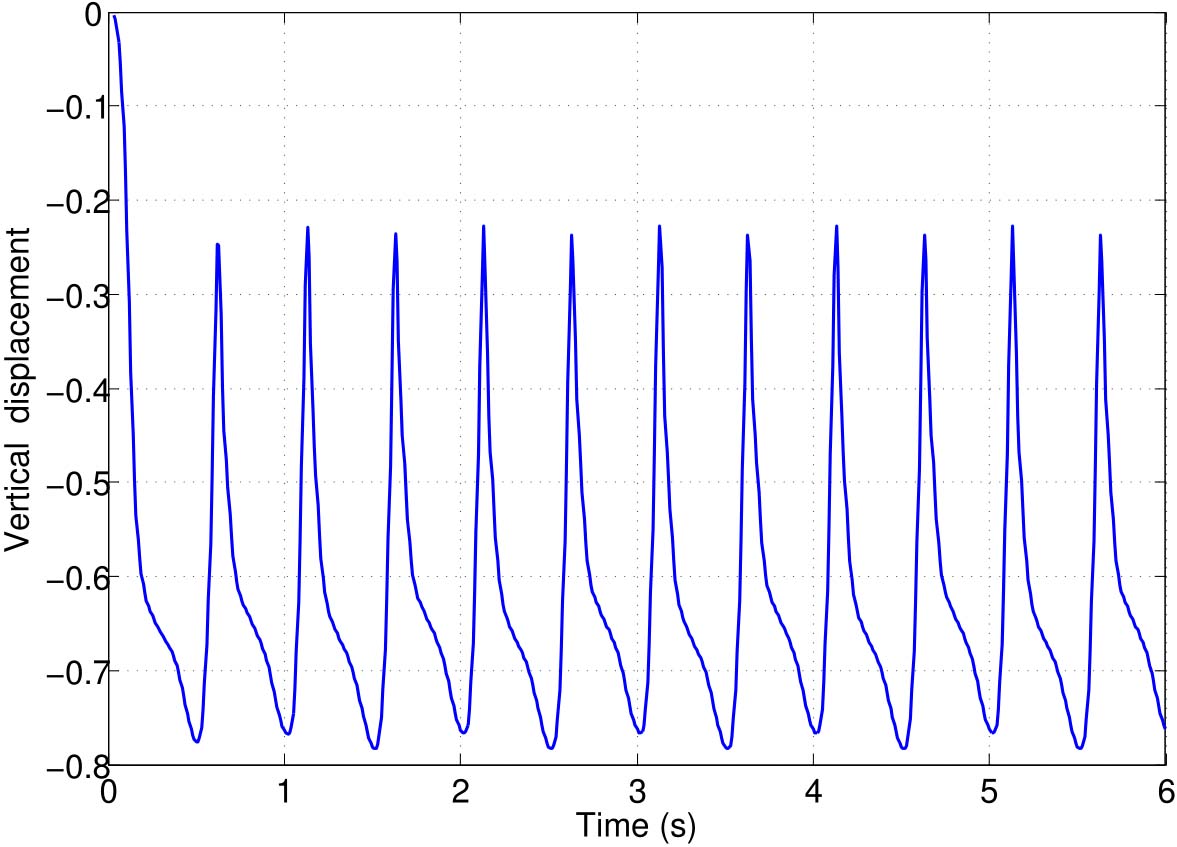}
	\end{minipage}   
	\captionsetup{justification=centering}
	\caption {\scriptsize Evolution of horizontal and vertical displacement at top right corner of the leaflet.} 
	\label{Evolution of horizontal and vertical displacement at top right corner of the leaflet}
\end{figure}

Previously published numerical results are qualitatively similar to those in Figure \ref{Configuration of leaflet and magnitude of velocity on the adaptive fluid mesh} but show some quantitative variations. For example, \cite{baaijens2001fictitious} solved a fully-coupled system but the coupling is limited to a line, and the solid in their results (Figure 7 (l)) behaves as if it is slightly harder. Alternatively, \cite{Yu_2005} used a fractional step scheme to solve the FSI equations combined with a penalty method to enforce the incompressibility condition. In their results (Fig. 3 (h)) the leaflet behaves as if it is slightly softer than \cite{baaijens2001fictitious} and harder than \cite{Kadapa_2016}. In \cite{Kadapa_2016} a beam formulation is used to describe the solid. The fluid mesh is locally refined using hierarchical B-Splines, and the FSI equation is solved monolithically. The leaflet in their results (Fig. 34) behaves as softer than the other two considered here. Our results in Figure \ref{Configuration of leaflet and magnitude of velocity on the adaptive fluid mesh} are most similar to those of \cite{Kadapa_2016}. This may be seen more precisely by inspection of the graphs of the oscillatory motion of the leaflet tip in Figure \ref{Evolution of horizontal and vertical displacement at top right corner of the leaflet}, corresponding to Fig. 32 in \cite{Kadapa_2016}. We point out here that Taylor-Galerkin method has also been used to solve the convection step for this test, and we gain almost the same accuracy using the same time step $\Delta t=5.0\times 10^{-4}s$. Having validated our results for this example against the work of others, we shall use this test case to further explore more details of our method. 

\begin{figure}[h!]
	\centering
	\includegraphics[width=3.5in,angle=0]{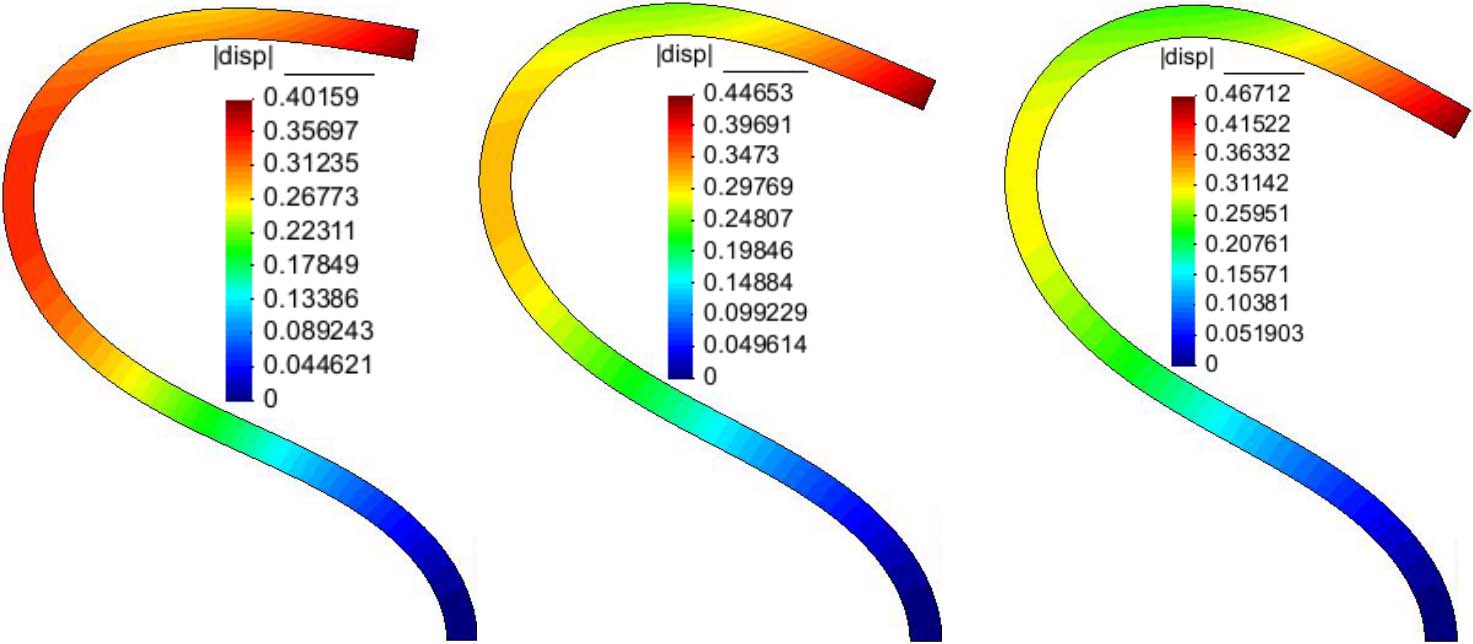}
	\captionsetup{justification=centering}
	\caption*{\scriptsize (a)coarse \qquad\qquad\qquad\qquad  (b)medium \qquad\qquad\qquad\qquad (c)fine}	
	\caption {\scriptsize Configuration of leaflet for different mesh ratio $r_m$, \\
		and contour plots of displacement magnitude at $t=0.6s$.} 
	\label{mesh ratio test}
\end{figure}

We commence by testing the influence of the ratio of fluid and solid mesh sizes $r_m$=(local fluid element area)/(solid element area). Fixing the fluid mesh size, three different solid mesh sizes are chosen: coarse (640 linear triangles with 403 nodes $r_m\approx1.5$), medium (1200 linear triangles with 794 nodes $r_m\approx3.0$) and fine (2560 linear triangles with 1445 nodes $r_m\approx5.0$), and a stable time step $\Delta t=5.0\times 10^{-4}s$ is used. From these tests we observe that there is a slight difference in the solid configuration for different meshes, as illustrated at $t=0.6s$ in Figure \ref*{mesh ratio test}. Significantly however, the difference in displacement decreases as the solid mesh becomes finer. Further, we found that $1.5\leq r_m\leq5.0$ ensures the stability of the proposed approach. Note that we use a 9-node quadrilateral for the fluid velocity and 3-node triangle for solid velocity, so $r_m\approx3.0$ means the fluid and solid mesh locally have a similar number of nodes for velocity.

\begin{figure}[h!]
	\begin{minipage}[t]{0.5\linewidth}
		\centering  
		\includegraphics[width=2in,angle=0]{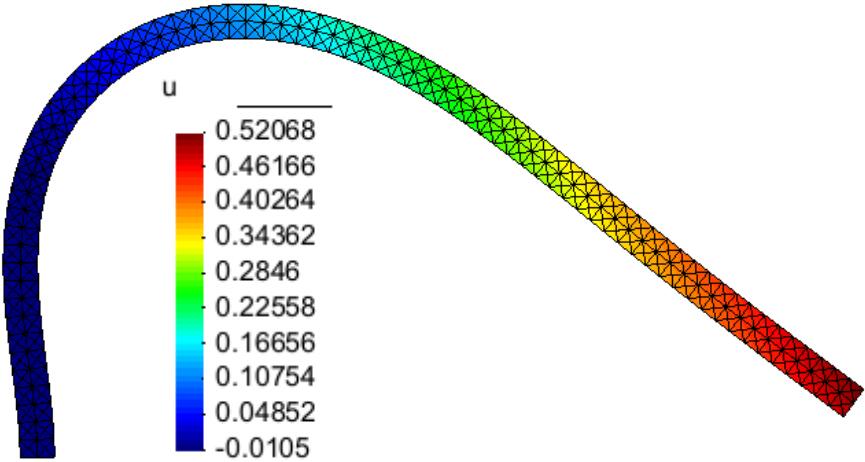}	
		\caption*{\scriptsize(a) Coarse}
	\end{minipage}
	\begin{minipage}[t]{0.5\linewidth}
		\centering  
		\includegraphics[width=2in,angle=0]{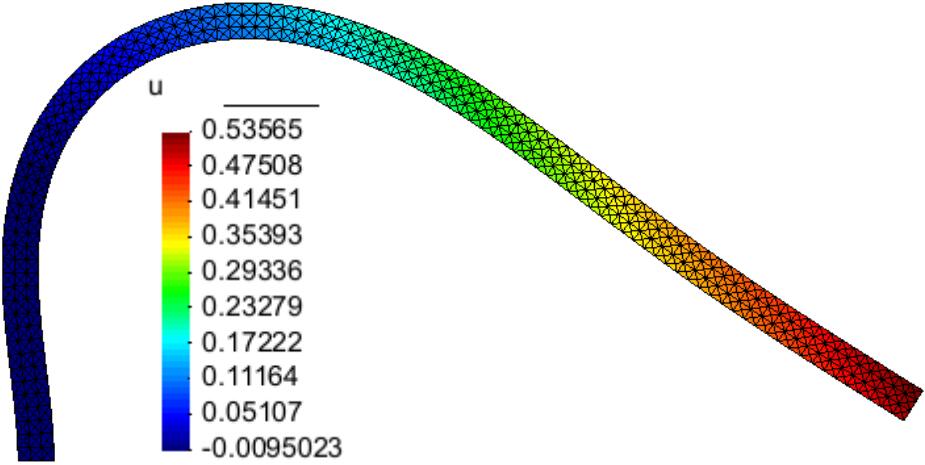}
		\caption*{\scriptsize(b) Medium}
	\end{minipage}   
	\begin{minipage}[t]{0.5\linewidth}
		\centering  
		\includegraphics[width=2in,angle=0]{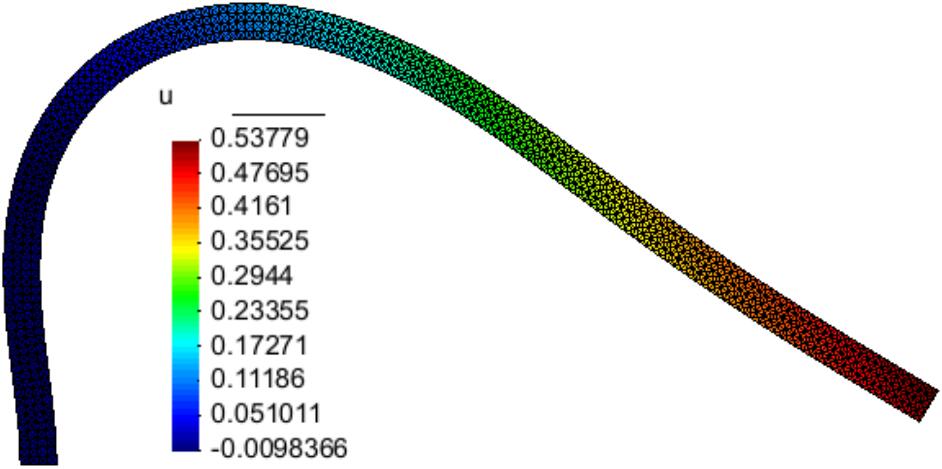}
        \caption*{\scriptsize(c) Fine}
	\end{minipage}
	\begin{minipage}[t]{0.5\linewidth}
		\centering  
		\includegraphics[width=2in,angle=0]{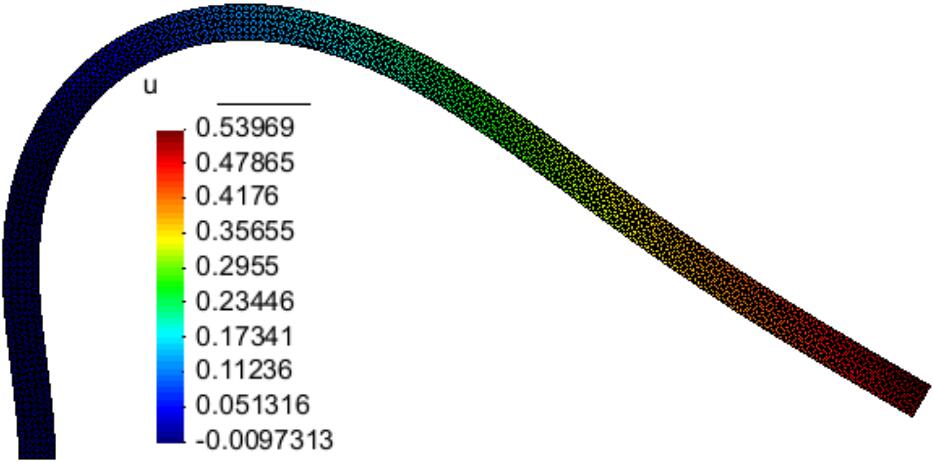}
		\caption*{\scriptsize(d) Very fine}
	\end{minipage}   		
	\captionsetup{justification=centering}
	\caption {\scriptsize Contour plots of horizontal velocity at $t=0.5s$.} 
	\label{mesh convergence test}
\end{figure}

We next consider convergence tests undertaken for refinement of both the fluid and solid meshes with the fixed ratio of mesh sizes $r_m\approx3.0$. Four different levels of meshes are used, the solid meshes are: coarse (584 linear triangles with 386 nodes), medium (1200 linear triangles with 794 nodes), fine (2560 linear triangles with 1445 nodes), and very fine (3780 linear triangles with 2085 nodes). The fluid meshes have the corresponding sizes with the solid at their maximum refinement level. As can be seen in Figure \ref{mesh convergence test} and Table \ref{Comparison of maximum velocity for different meshes}, the velocity is converging as the mesh becomes finer.

\begin{table}[h!]
\newcommand{\tabincell}[2]{\begin{tabular}{@{}#1@{}}#2\end{tabular}}	
	\centering
	\begin{tabular}{|c|c|}
		\hline
		{Between different mesh sizes}  & \tabincell{c} {Difference of maximum \\  horizontal velocity at $t=0.5s$} \\
		\hline
		coarse and medium & 0.01497\\
		\hline
		medium and fine & 0.00214 \\
		\hline
		fine and very fine & 0.00190 \\
		\hline
	\end{tabular}
	\caption{Comparison of maximum velocity for different meshes.}
	\label{Comparison of maximum velocity for different meshes}
\end{table}

\begin{figure}[h!]
	\begin{minipage}[t]{0.5\linewidth}
		\centering  
		\includegraphics[width=2in,angle=0]{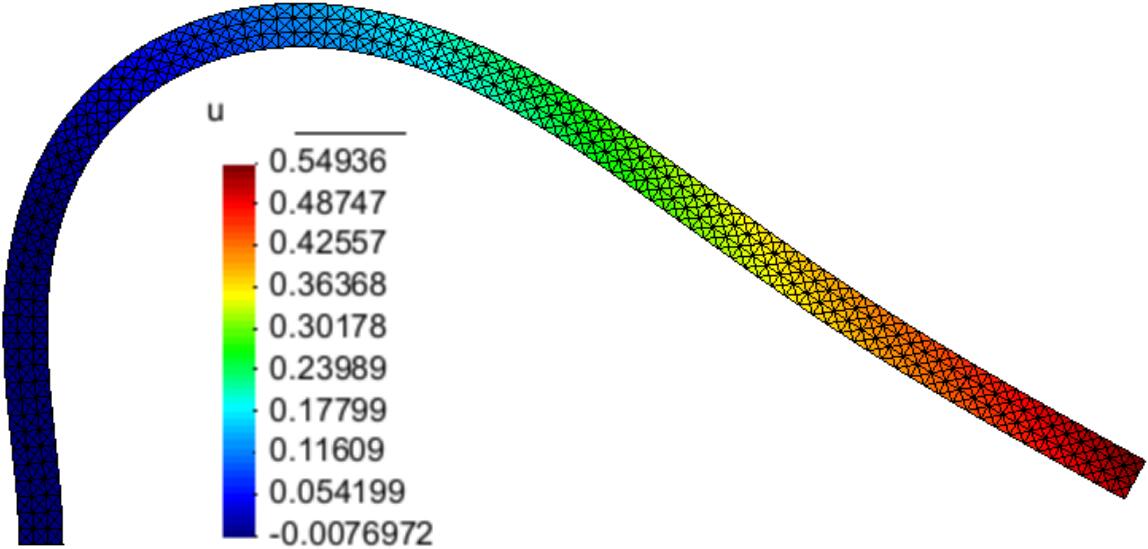}	
        \captionsetup{justification=centering}
		\caption*{\scriptsize(a) $\Delta t=2.0\times 10^{-3}s$ \\ (breaks down at $t=0.61s$).}
	\end{minipage}
	\begin{minipage}[t]{0.5\linewidth}
		\centering  
		\includegraphics[width=2in,angle=0]{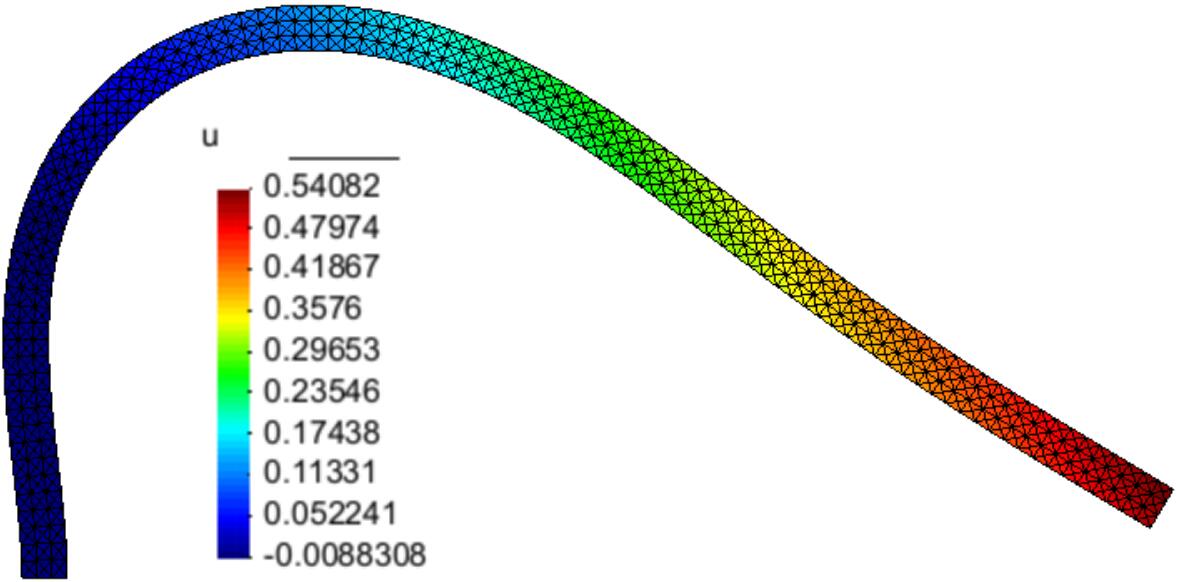}
		\caption*{\scriptsize(b) $\Delta t=1.0\times 10^{-3}s$.}
	\end{minipage}   
	\begin{minipage}[t]{0.5\linewidth}
		\centering  
		\includegraphics[width=2in,angle=0]{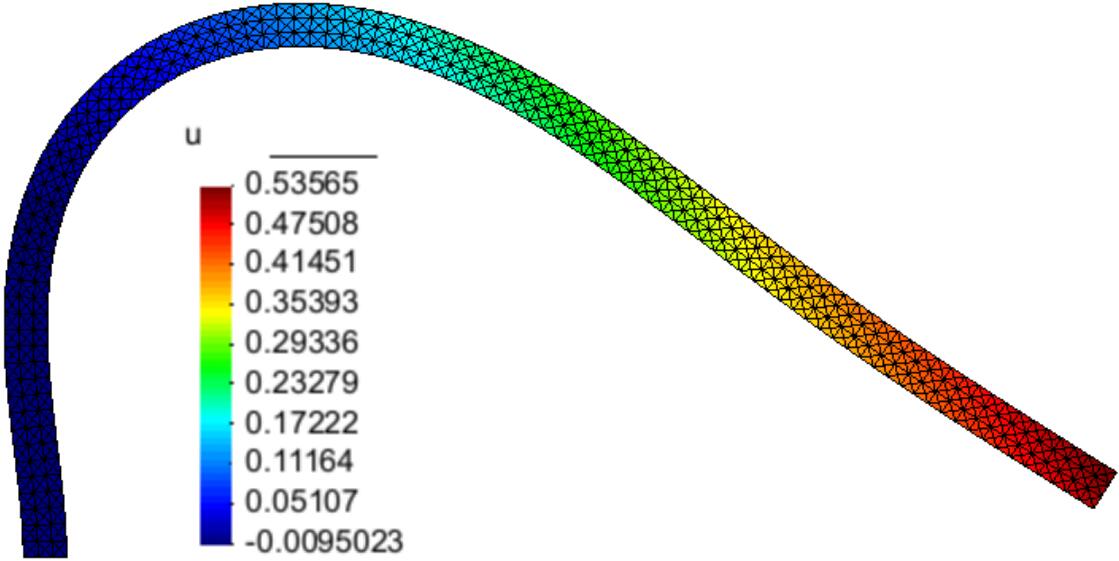}
		\caption*{\scriptsize(c) $\Delta t=5.0\times 10^{-4}s$.}
	\end{minipage}
	\begin{minipage}[t]{0.5\linewidth}
		\centering  
		\includegraphics[width=2in,angle=0]{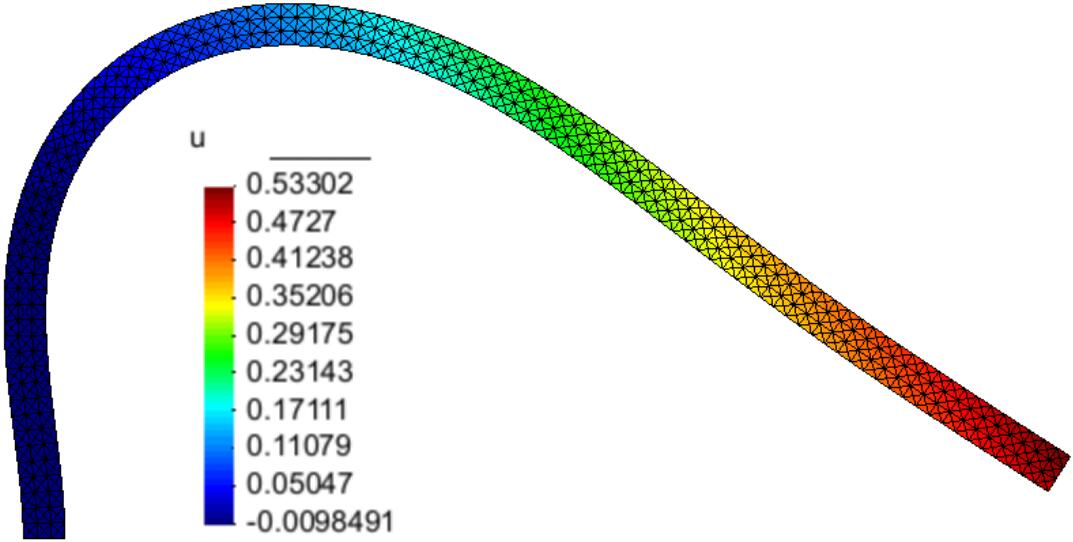}
		\caption*{\scriptsize(d) $\Delta t=2.5\times 10^{-4}s$.}
	\end{minipage}   		
	\captionsetup{justification=centering}
	\caption {\scriptsize Contour plots of horizontal velocity at $t=0.5s$.} 
	\label{time convergence test}
\end{figure}

In addition, we consider tests of convergence in time using a fixed ratio of fluid and solid mesh sizes $r_m\approx3.0$. Using the medium solid mesh size and the same fluid mesh size as above, results are shown in Figure \ref{time convergence test} and Table \ref{Comparison of maximum velocity for different time step size}. From these it can be seen that the velocities are converging as the time step decreases.

\begin{table}[h!]
	\newcommand{\tabincell}[2]{\begin{tabular}{@{}#1@{}}#2\end{tabular}}	
	\centering
	\begin{tabular}{|c|c|}
		\hline
		{Steps sizes compared}  & \tabincell{c} {Difference of maximum \\  horizontal velocity at $t=0.5s$} \\
		\hline
		$\Delta t=2.0\times 10^{-3}$ and $\Delta t=1.0\times 10^{-3}$ & 0.00854\\
		\hline
		$\Delta t=1.0\times 10^{-3}$ and $\Delta t=5.0\times 10^{-4}$ & 0.00517 \\
		\hline
		$\Delta t=5.0\times 10^{-4}$ and $\Delta t=2.5\times 10^{-4}$ & 0.00263 \\
		\hline
	\end{tabular}
	\caption{Comparison of maximum velocity for different time step size.}
	\label{Comparison of maximum velocity for different time step size}
\end{table}

\begin{figure}[h!]
	\begin{minipage}[t]{0.5\linewidth}
		\centering  
		\includegraphics[width=1.5in,angle=0]{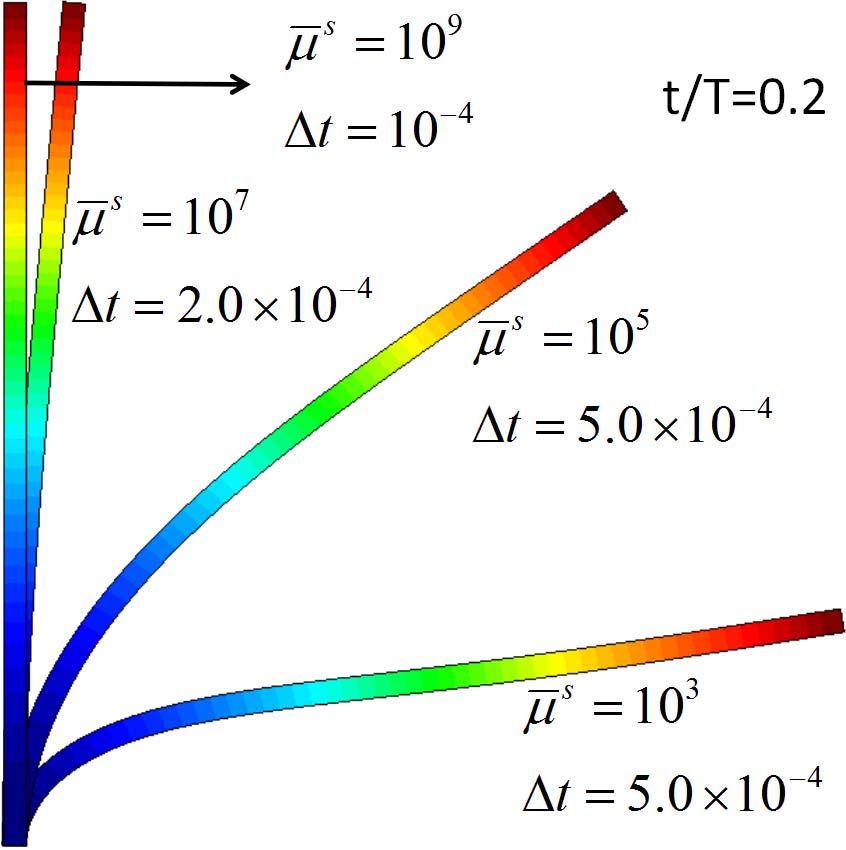}	
		\captionsetup{justification=centering}
		\caption*{\scriptsize(a) $\rho^r=1, Re=100$ and $Fr=0$.}
	\end{minipage}
	\begin{minipage}[t]{0.5\linewidth}
		\centering  
		\includegraphics[width=2.5in,angle=0]{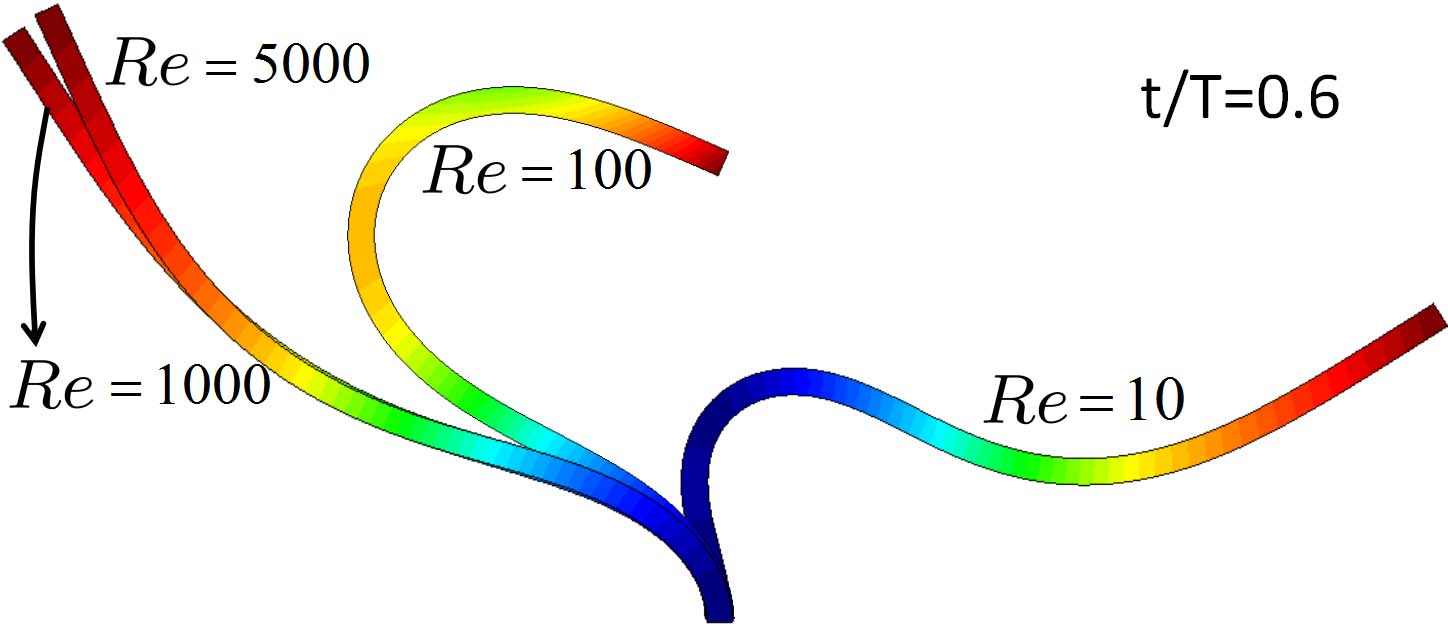}
		\caption*{\scriptsize(b) $\rho^r=1, \bar{\mu}^s=10^3$ and $Fr=0$.}
	\end{minipage}   
	\begin{minipage}[t]{0.5\linewidth}
		\centering  
		\includegraphics[width=1.5in,angle=0]{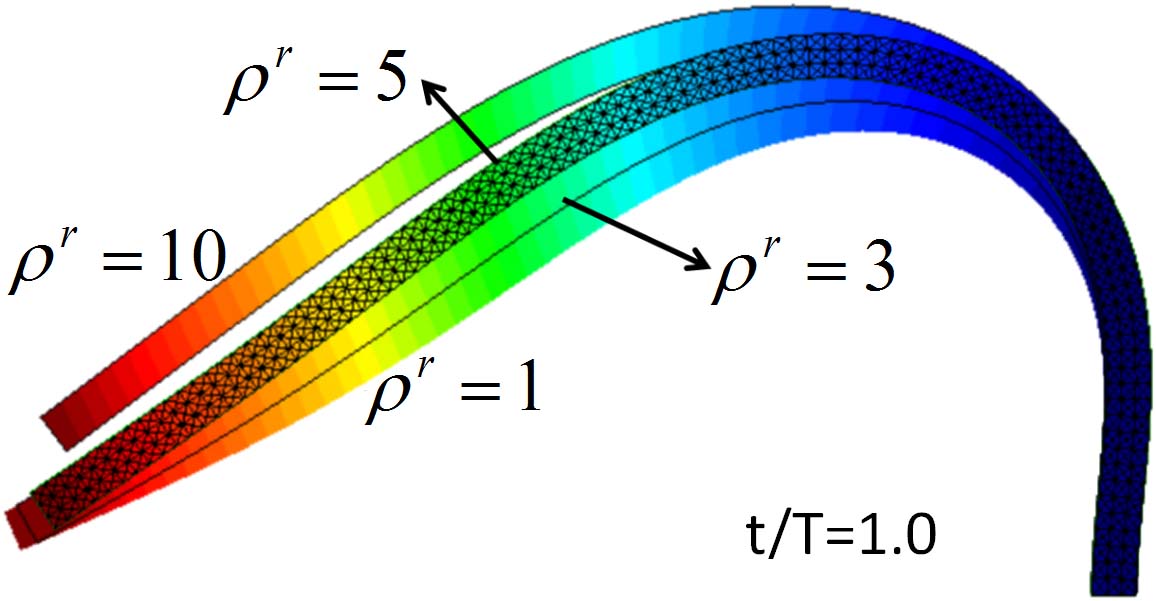}
		\caption*{\scriptsize(c) $Re=100, \bar{\mu}^s=10^3$ and $Fr=0$.}
	\end{minipage}
	\begin{minipage}[t]{0.5\linewidth}
		\centering  
		\includegraphics[width=1.5in,angle=0]{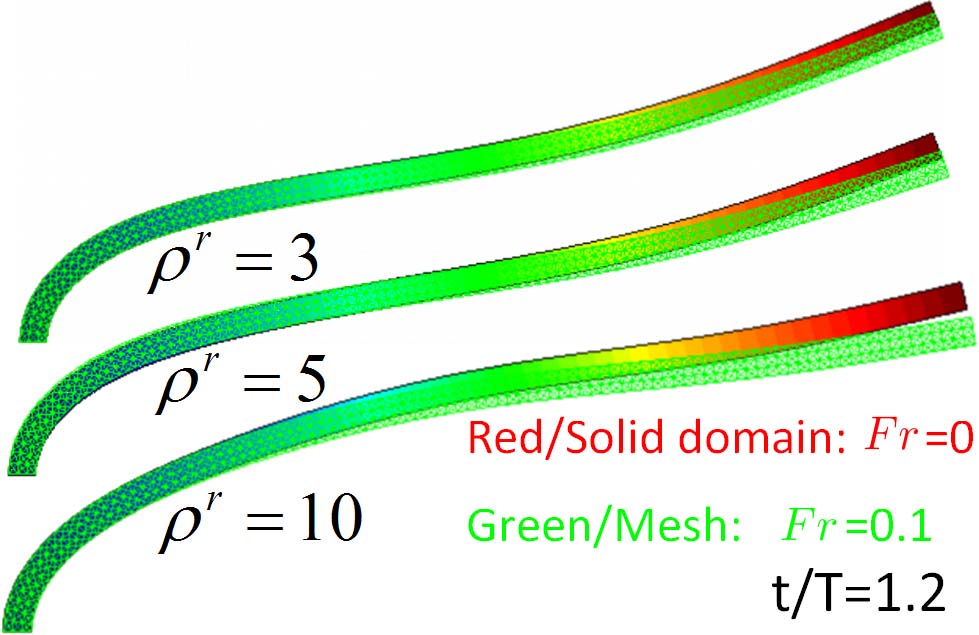}
		\caption*{\scriptsize(d) $Re=100$ and $\bar{\mu}^s=10^3$.}
	\end{minipage}   		
	\captionsetup{justification=centering}
	\caption {\scriptsize Parameters sets and results, $\Delta t=5.0\times 10^{-4}s$ for Group (b)$\sim$(d).} 
	\label{permutation of parameters}
\end{figure}

Finally, in order to assess the robustness of our approach, we vary each of the physical parameters using three different cases as shown in Figure \ref{permutation of parameters}. A medium mesh size with fixed $r_m\approx3.0$ is used to undertake all of these tests. The dimensionless parameters shown in Figure \ref{permutation of parameters} are defined as: $\rho^r=\frac{\rho^s}{\rho^f}, \bar{\mu}^s=\frac{\mu^s}{\rho^fU^2}, Re=\frac{\rho^fUH}{\mu^f}$ and $Fr=\frac{gH}{U^2}$ where the average velocity $U=10$ in this example. The period of inlet flow is $T=1$.

It can be seen from the results of group (a) that the larger the value of shear modulus $\bar{\mu}^s$ the harder the solid behaves, however a smaller time step is required. For the case of $\bar{\mu}^s=10^9$, the solid behaves almost like a rigid body, as we would expect. From the results of group (b) it is clear that the Reynolds Number $\left(Re\right)$ has a large influence on the behavior of the solid. The density and gravity have relatively less influence on the behavior of solid in this problem which can be seen from the results of group (c) and group (d) respectively.

\subsection{Oscillating disc surrounded by fluid}
This example is taken from \cite{Zhao_2008} and used to validate the conservation of mass and energy using the proposed method. The computational domain is a square $[0,1]\times[0,1]$ with a homogeneous Dirichlet boundary conditions imposed for velocity, whilst the pressure is fixed to be zero at the left-bottom corner of the square. A soft solid disc is initially located in the middle of the square and has a radius of $0.2$. The initial velocity of the fluid and solid are prescribed by the following stream function
\begin{equation*}
	\Psi=\Psi_0{\rm sin}(ax){\rm sin}(by),
\end{equation*}
where $\Psi_0=5.0\times10^{-2}$ and $a=b=2\pi$. The whole system then evolves from this initial condition.

\begin{figure}[h!]
	\begin{minipage}[h!]{0.5\linewidth}
		\centering  
		\includegraphics[width=2.5in,angle=0]{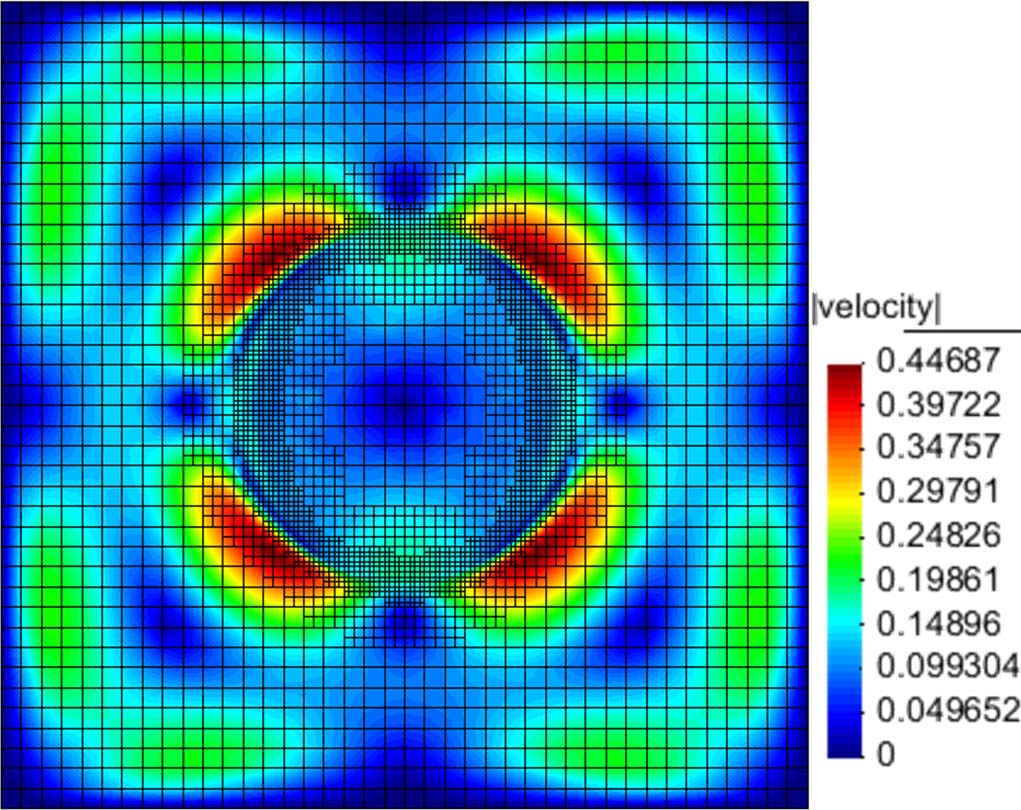}	
		\caption*{\scriptsize(a) Velocity norm.}
	\end{minipage}
	\begin{minipage}[h!]{0.5\linewidth}
		\centering  
		\includegraphics[width=2in,angle=0]{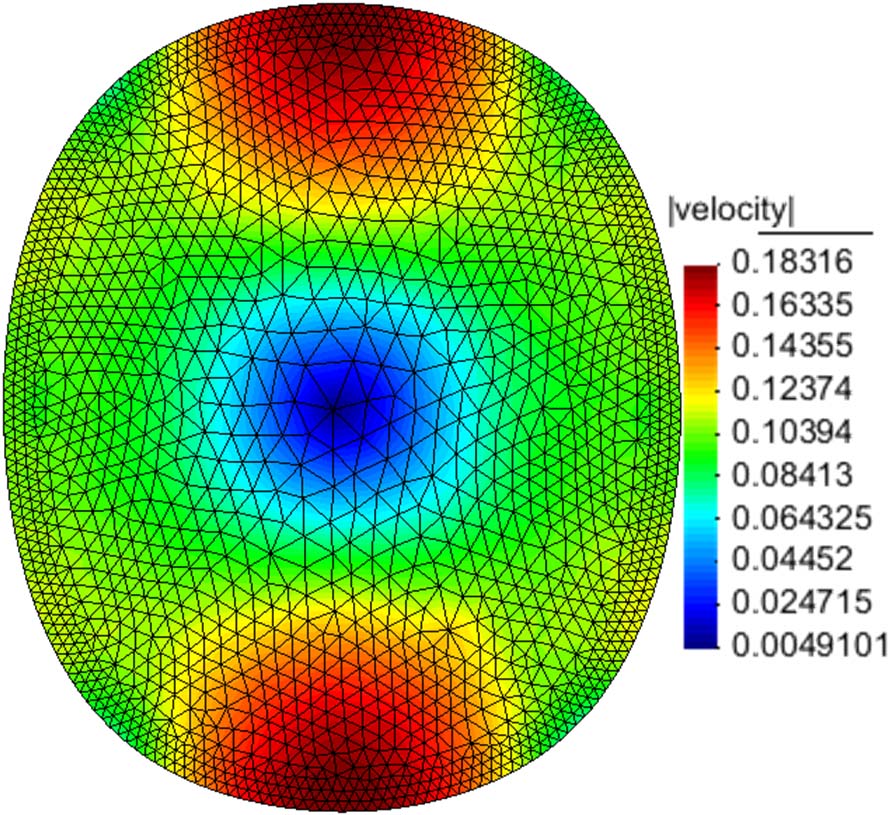}
		\caption*{\scriptsize(b) Defomation of solid.}		
	\end{minipage}     		
	\captionsetup{justification=centering}
	\caption {Velocity norm and solid defomation at $t=0.5$ \\ for $\rho^s=1.0$, $\Delta t=10^{-3}$ (medium mesh).} 
	\label{Velocity norm on the adpative mesh and defomation of solid}	
\end{figure}

\begin{figure}[h!]
	\begin{minipage}[h!]{0.5\linewidth}
		\centering  
		\includegraphics[width=2.4in,angle=0]{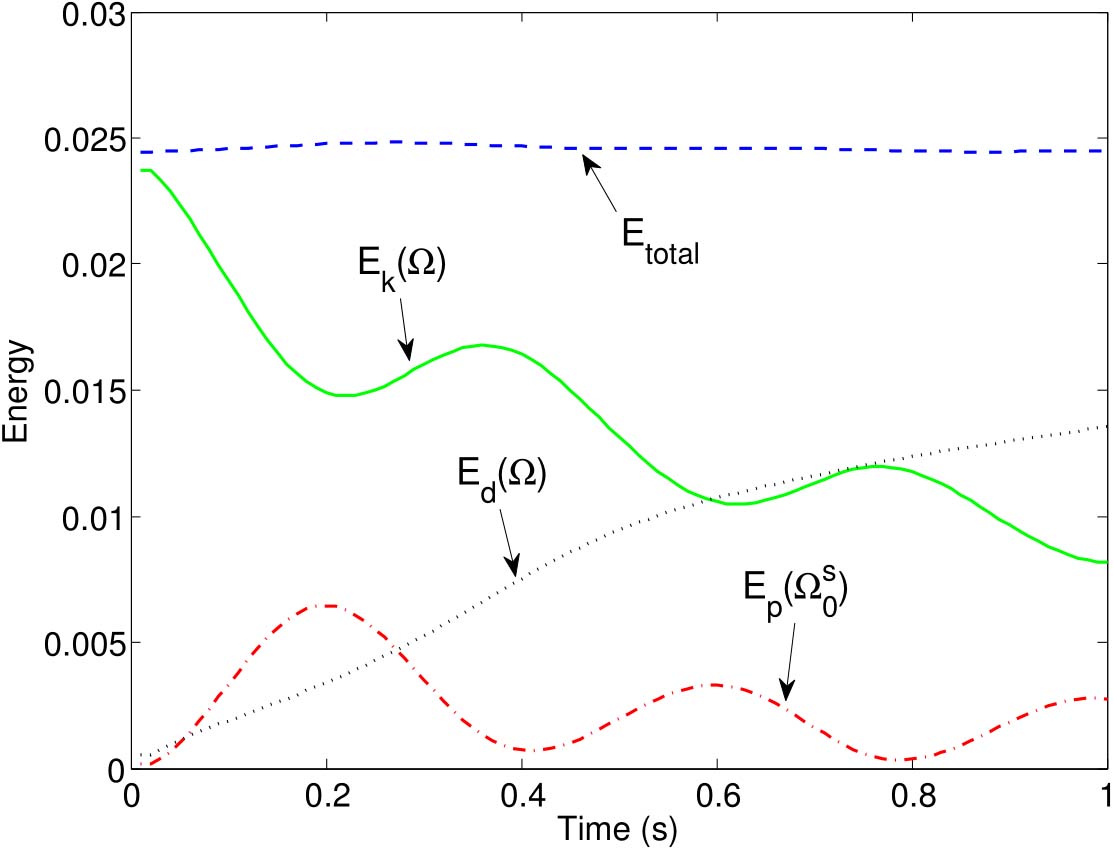}	
		\caption*{\scriptsize(a)  $\rho^f=\rho^s=1.0$.}
	\end{minipage}
	\begin{minipage}[h!]{0.5\linewidth}
		\centering  
		\includegraphics[width=2.4in,angle=0]{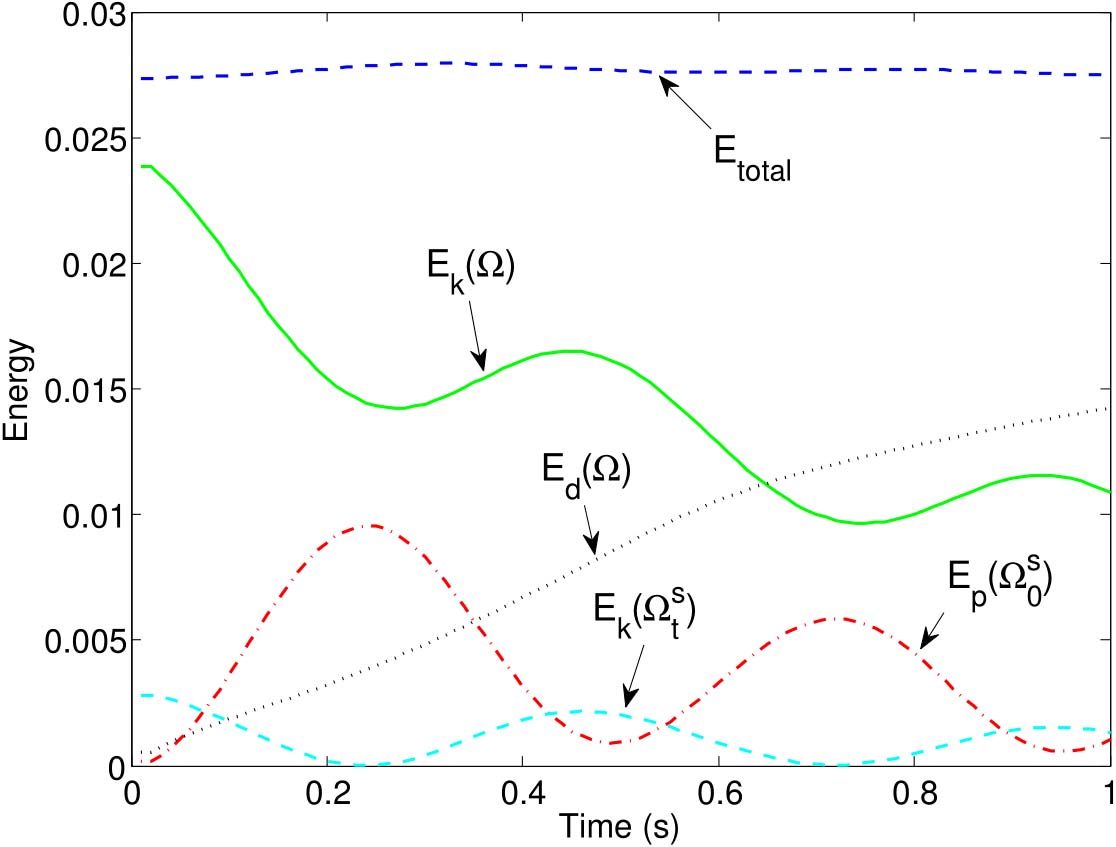}
		\caption*{\scriptsize(b)  $\rho^f=1.0$, $\rho^s=2.0$.}		
	\end{minipage}     		
	\captionsetup{justification=centering}
	\caption {\scriptsize Energy evolution of an oscillating disc, $\Delta t=10^{-3}$ (medium mesh).} 
	\label{Energy evolution of the oscillating disc}		
\end{figure}

We first use the same parameters as used in \cite{Zhao_2008}: $\rho^f=\rho^s=1.0$, $\mu^f=10^{-3}$ and $\mu^s=1.0$, then we swith to $\rho^s=2.0$ and $\rho^s=10$, and undertake the tests with the other parameters unchanged. Three different initial meshes for the square (coarse: $20\times 20$, medium: $40\times 40$ and fine: $80\times 80$) and 2-level adaptive refinement near the interface based on these initial meshes are used. The solid mesh has a similar number of nodes with the fluid mesh near the interface. A snapshot of the velocity norm on the medium adpative mesh, and the corresponding deformation of the disc at $t=0.5$, are plotted in Figure \ref{Velocity norm on the adpative mesh and defomation of solid}.
The energy of this FSI system is computed as follows.\\
Kinetic energy in $\Omega$:
\begin{equation}
	E_k(\Omega)=\frac{\rho^f}{2}\int_{\Omega}\left|{\bf u}\right|^2.
\end{equation}
Kinetic energy in $\Omega_t^s$:
\begin{equation}
	E_k(\Omega_t^s)=\frac{\rho^s-\rho^f}{2}\int_{\Omega_t^s}\left|{\bf u}\right|^2.
\end{equation}
Viscous dissipation in $\Omega$:
\begin{equation}
	E_d(\Omega)=\int_0^t\int_{\Omega}{\tau_{ij}^f}{\frac{\partial u_i}{\partial x_j}}.
\end{equation}
Potential energy of solid:
\begin{equation}\label{solid pontential equattion}
	E_p(\Omega_0^s)
	=\frac{\mu^s}{2}\int_{\Omega_0^s}\left(tr_{{\bf F}{\bf F}^{\rm T}}-d\right).
\end{equation}
Analytically, the total energy 
\begin{equation}
	E=E_k(\Omega)+E_k(\Omega_t^s)+E_d(\Omega)+E_p(\Omega_0^s)
\end{equation}
should be a constant. This is considered in the plots of Figure \ref{Energy evolution of the oscillating disc}. When the fluid and solid have the same density, the maximum variation of total energy is around $1.6\%$ ($t=0.26$), as shown Figure \ref{Energy evolution of the oscillating disc} (a), and when their densities are different ($\rho^s=2.0$), as shown in \ref{Energy evolution of the oscillating disc} (b), the maximum variation of total energy is around $2.2\%$ at $t=0.31$. For the case of $\rho^s=10$ we have a similar result, with the maximum variation of total energy being about $4.9\%$ at $t=0.6$ using the same time step $\Delta t=10^{-3}$. 

\begin{figure}[h!]
	\begin{minipage}[h!]{0.5\linewidth}
		\centering  
		\includegraphics[width=2.4in,angle=0]{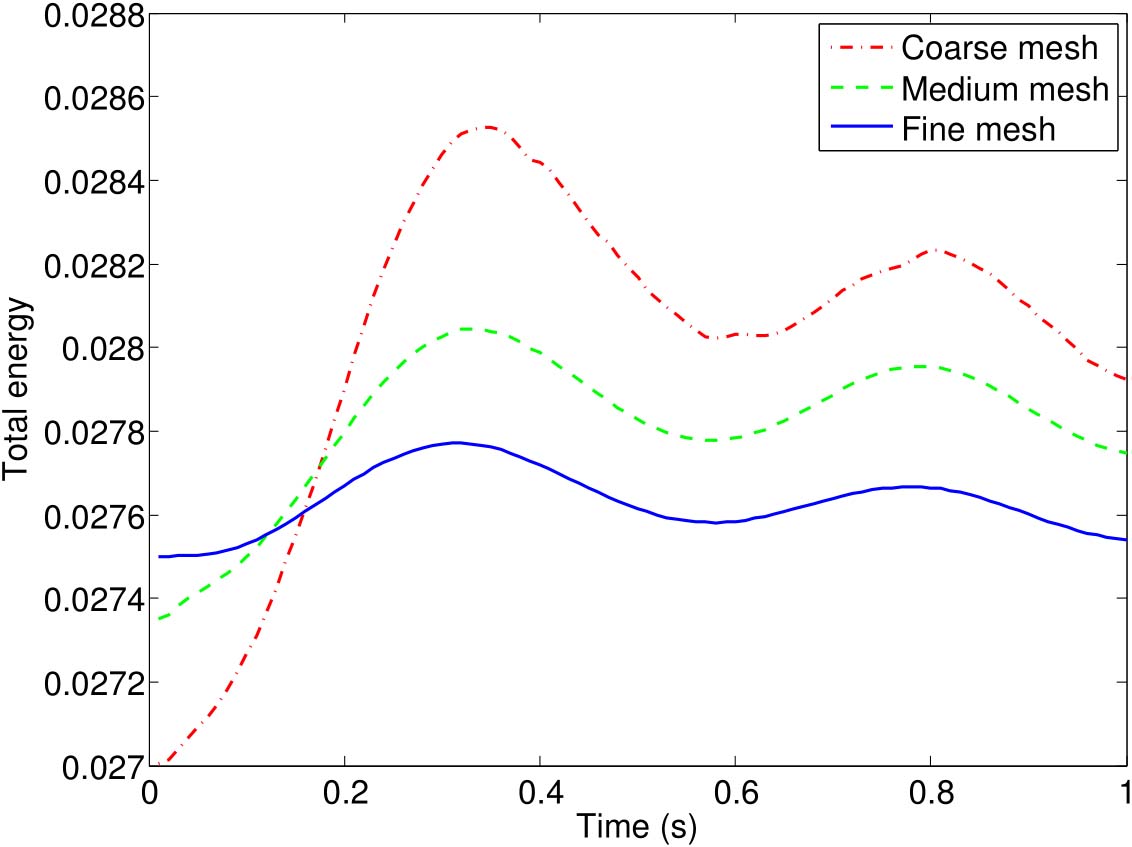}	
		\caption*{\scriptsize(a)  Engergy convergence.}
	\end{minipage}
	\begin{minipage}[h!]{0.5\linewidth}
		\centering  
		\includegraphics[width=2.25in,angle=0]{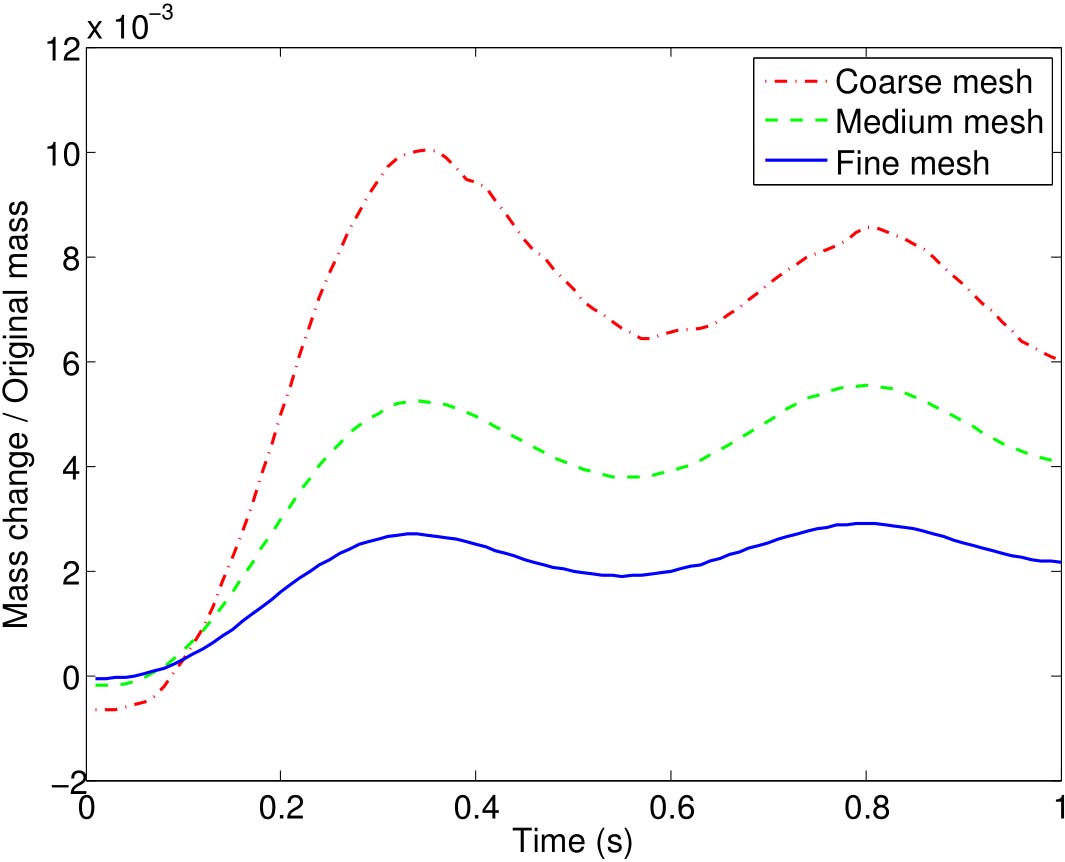}
		\caption*{\scriptsize(b)  Mass convergence.}		
	\end{minipage}     		
	\captionsetup{justification=centering}
	\caption {\scriptsize Convergence of energy and mass, $\rho^s=2.0$, $\Delta t=5.0\times 10^{-4}$.} 
	\label{Convergence of energy and mass}		
\end{figure}

We further verify the convergence of both energy and mass, which is clearly demonstrated in Figure \ref{Convergence of energy and mass}.

\subsection{Oscillation of a flexible leaflet oriented along the flow direction}
\begin{figure}[h!]
	\centering  
	\includegraphics[width=3in,angle=0]{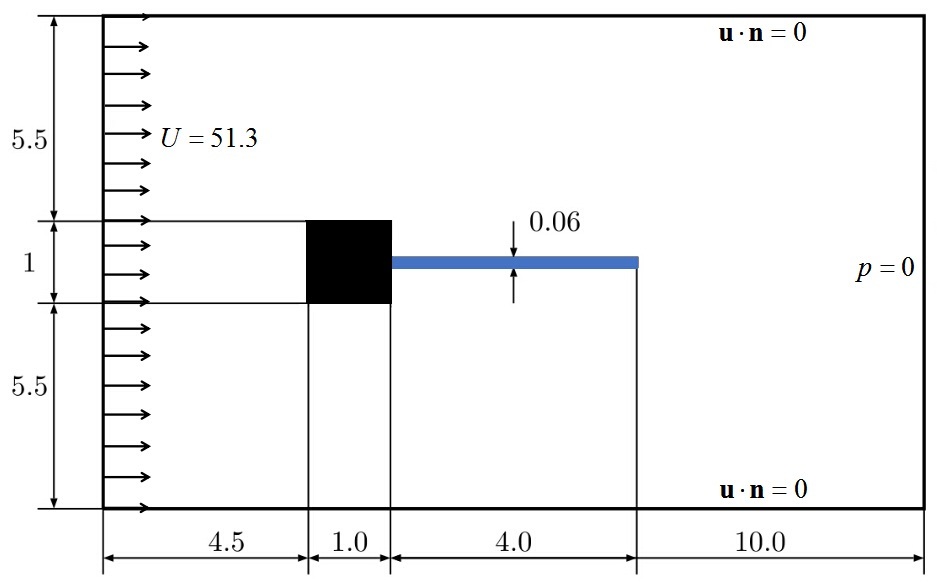}
	\captionsetup{justification=centering}
	\caption {\scriptsize Computational domain and boundary condition for oscillation of flexible leaflet.} 
	\label{Computational domain and boundary condition for oscillation of flexible leaflet}						
\end{figure}

The following test problem is taken from \cite{wall1999fluid}, which describes an implementation on an ALE fitted mesh. It has since been used as a benchmark to validate different numerical schemes \cite{Kadapa_2016,Hesch_2014}. The geometry and boundary conditions are shown in Figure \ref{Computational domain and boundary condition for oscillation of flexible leaflet}. 

\begin{figure}[h!]
	\begin{minipage}[t]{0.5\linewidth}
		\centering  
		\includegraphics[width=2.3in,angle=0]{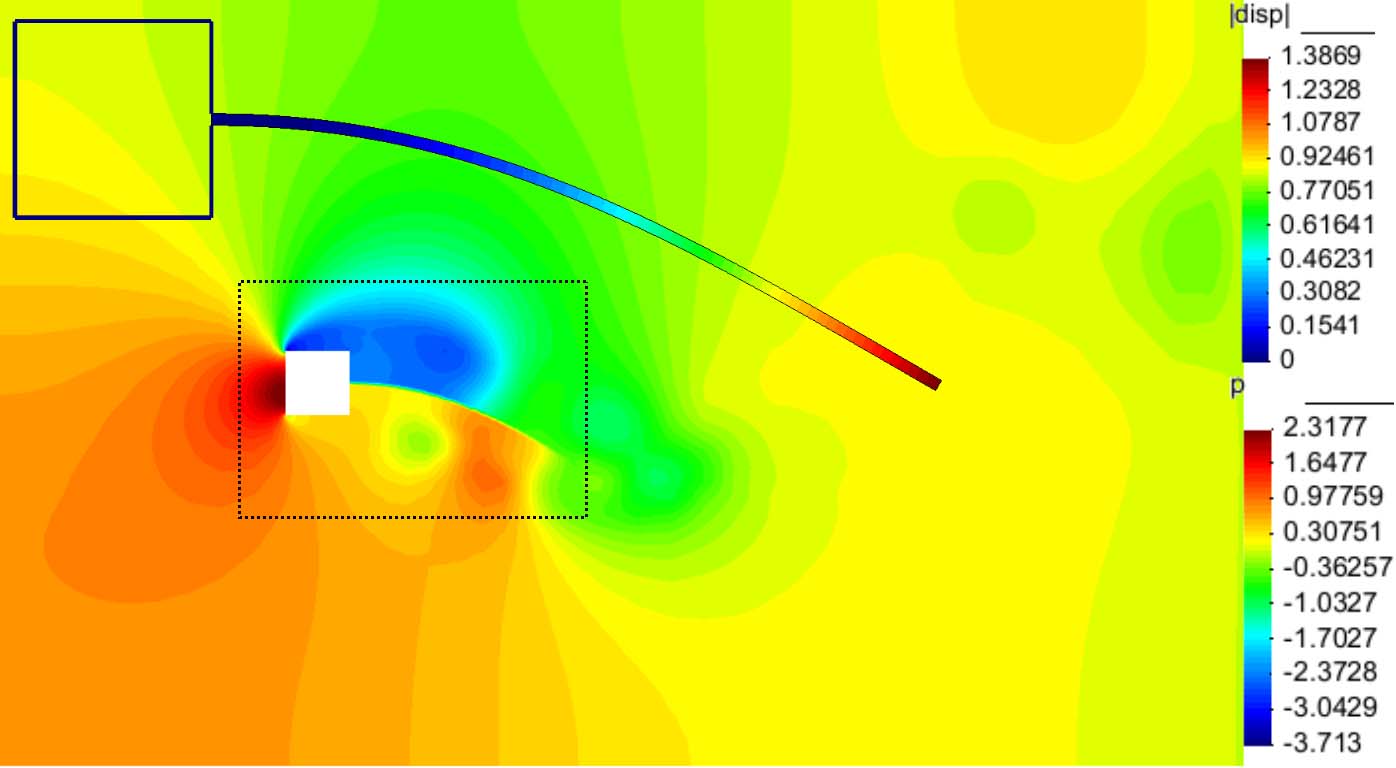}	
        \caption*{\scriptsize(a) Leaflet displacement and fluid pressure.}
	\end{minipage}
	\begin{minipage}[t]{0.5\linewidth}
		\centering  
		\includegraphics[width=2.3in,angle=0]{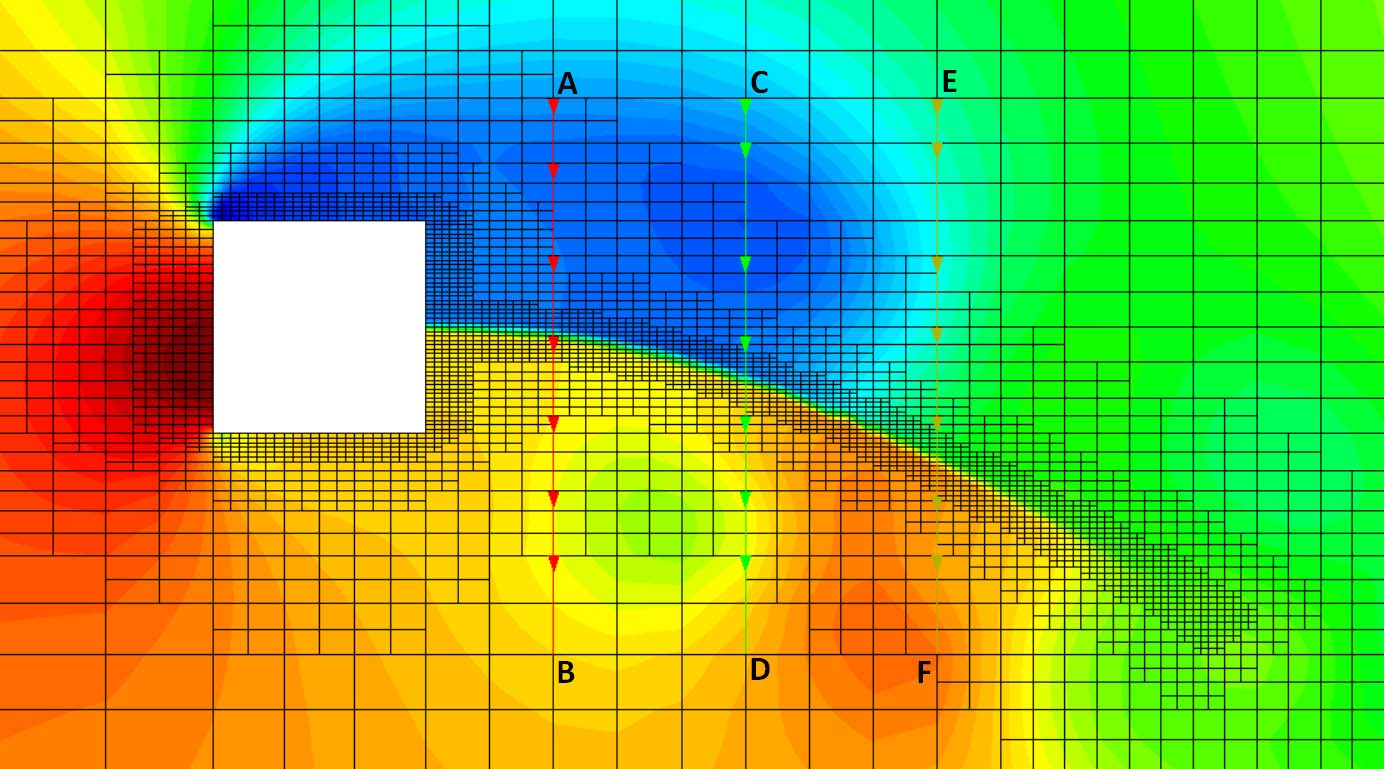}
        \caption*{\scriptsize(b) Mesh refinement near the structure.}		
	\end{minipage}     		
	\captionsetup{justification=centering}
	\caption {\scriptsize Contour plots of leaflet displacement and fluid pressure at $t=5.44s$.} 
	\label{Contours of leaflet displacement and fluid pressure}	
\end{figure}

\begin{figure}[h!]
	\centering  
	\includegraphics[width=3.0in,angle=0]{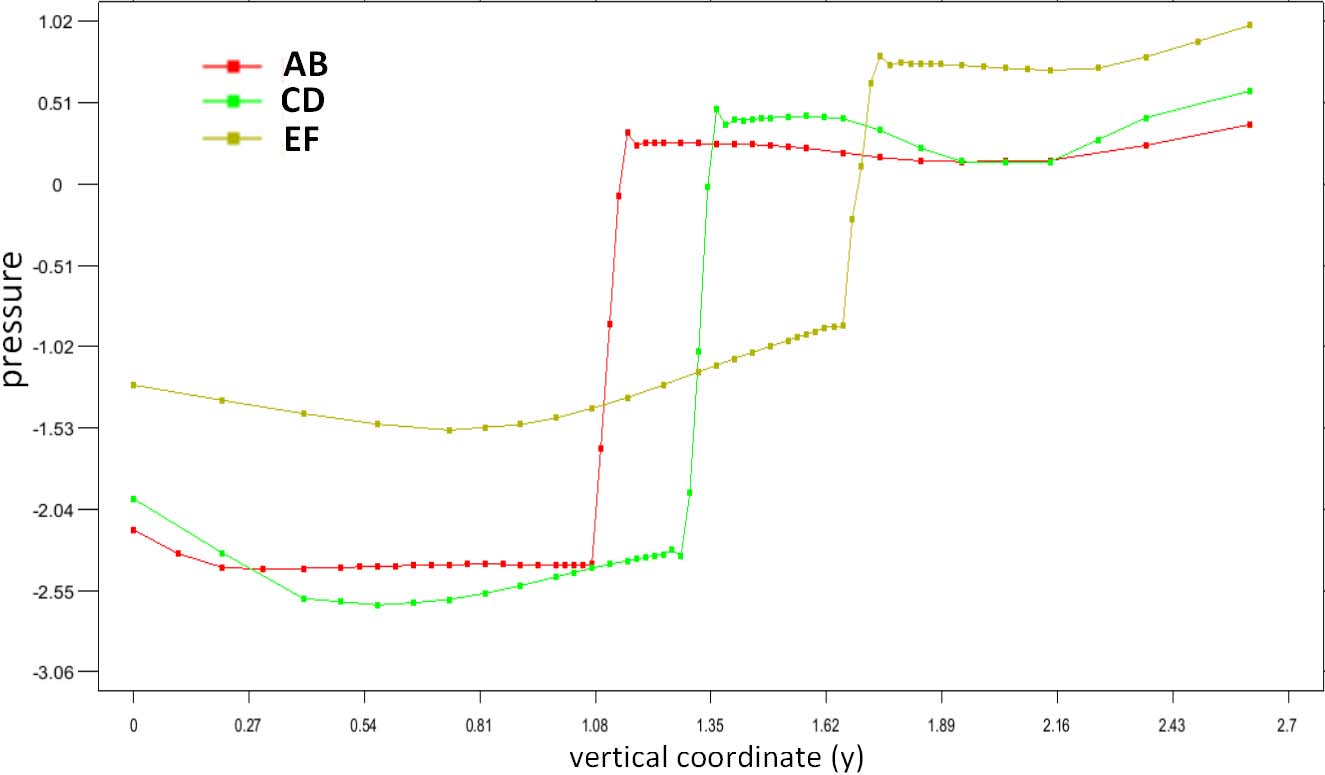}
	\captionsetup{justification=centering}
	\caption {\scriptsize Distribution of pressure across the leaflet on the three lines in Figure \ref{Contours of leaflet displacement and fluid pressure} (b).} 
	\label{Distribution of pressure across the leaflet}						
\end{figure}

For the fluid, the viscosity and density are $\mu^f=1.82\times 10^{-4}$ and $\rho^f=1.18\times 10^{-3}$ respectively. For the solid, we use shear modulus $\mu^s=9.2593\times 10^5$ and density $\rho^s=0.1$. The leaflet is discretized by 1063 3-node linear triangles with 666 nodes, and the corresponding fluid mesh locally has a similar node density to the leaflet ($r_m\approx3.0$). First the Least-squares method is tested and a stable time step $\Delta t=1.0\times 10^{-3}s$ is used. Snapshots of the leaflet deformation and fluid pressure at $t=5.44s$ are illustrated in Figure \ref{Contours of leaflet displacement and fluid pressure}. In Figure \ref{Distribution of pressure across the leaflet}, the distributions of pressure across the leaflet corresponding to the three lines (AB, CD and EF) in Figure \ref{Contours of leaflet displacement and fluid pressure} (b) are plotted, from which we can observe that the sharp jumps of pressure across the leaflet are captured.

\begin{figure}[h!]
	\begin{minipage}[t]{0.5\linewidth}
		\centering  
		\includegraphics[width=2.2in,angle=0]{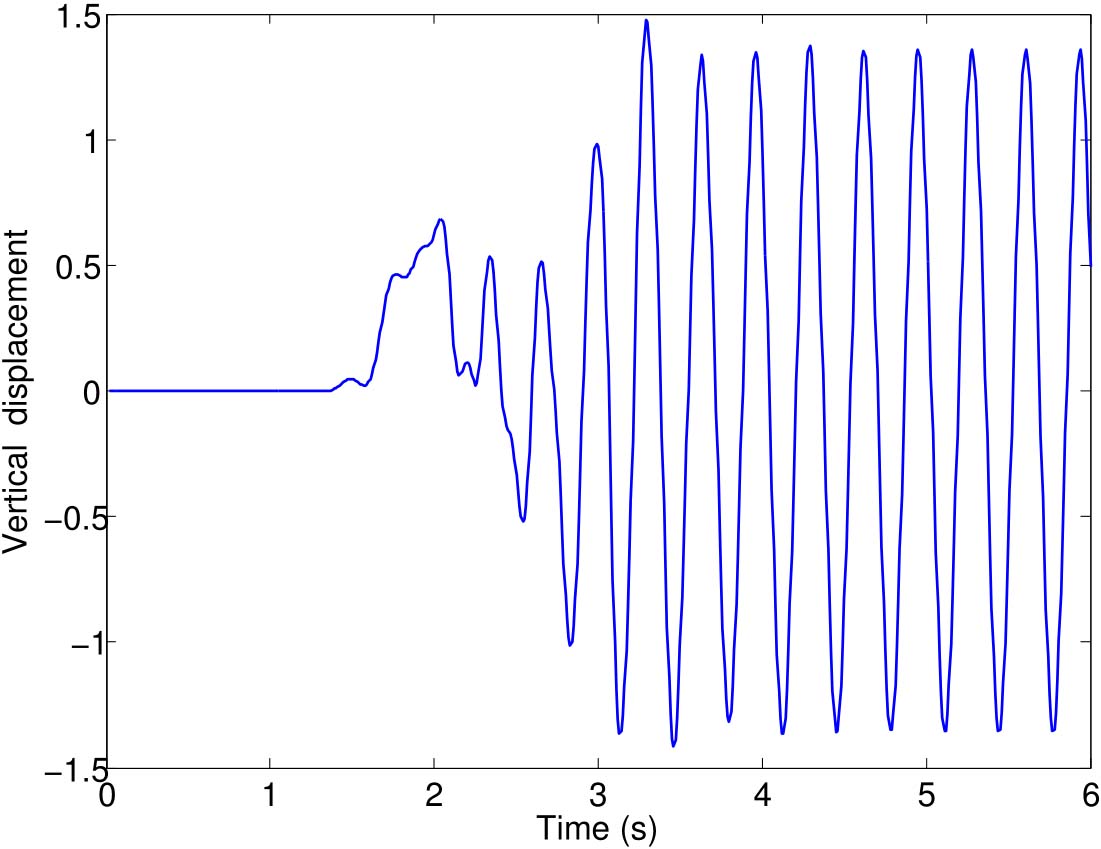}	
		\caption*{\scriptsize (a) Least-squares method.}
	\end{minipage}
	\begin{minipage}[t]{0.5\linewidth}
		\centering  
		\includegraphics[width=2.2in,angle=0]{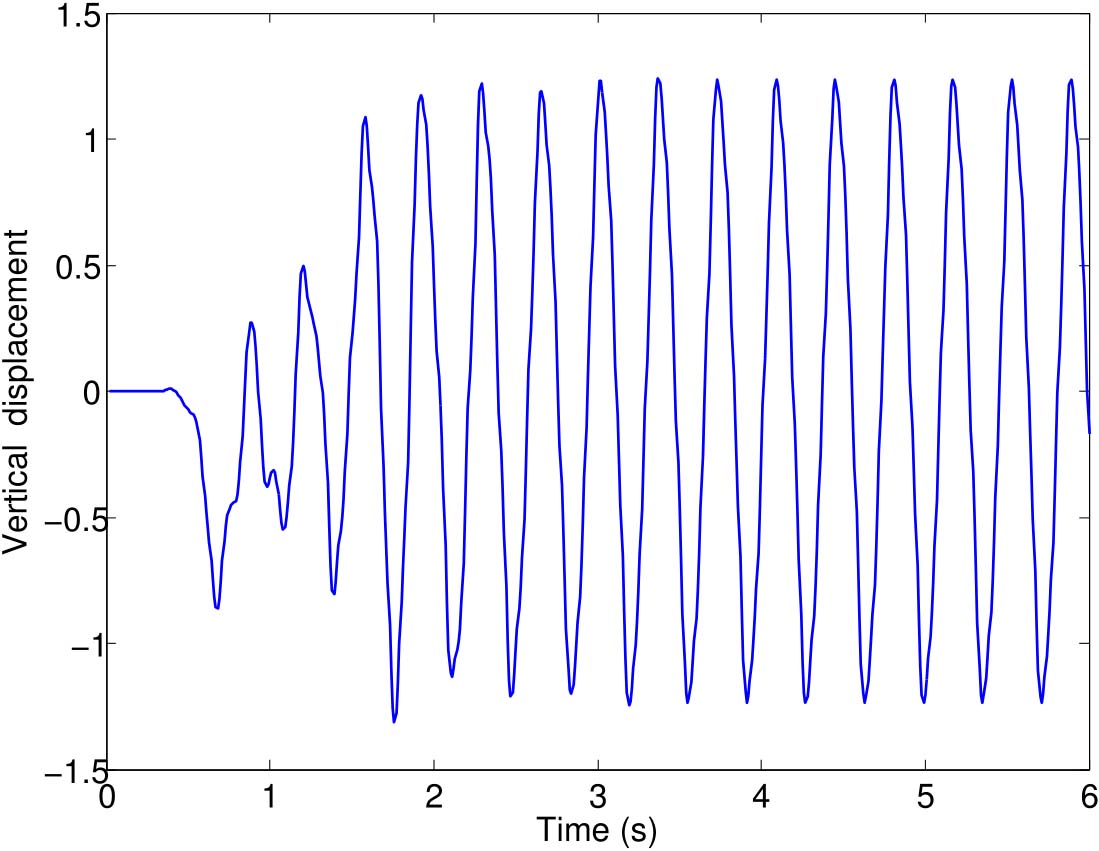}
		\caption*{\scriptsize (b) Taylor-Galerkin method.}		
	\end{minipage}     		
	\captionsetup{justification=centering}
	\caption {\scriptsize Displacement of leaflet tip as a function of time.} 
	\label{Displacement of leaflet tip as a function of time}	
\end{figure}

The evolution of the vertical displacement of the leaflet tip with respect to time is plotted in Figure \ref{Displacement of leaflet tip as a function of time}(a). Both the magnitude (1.34) and the frequency (2.94) have a good agreement with the result of \cite{wall1999fluid}, using a fitted ALE mesh and of \cite{Kadapa_2016}, using a monolithic unfitted mesh approach. Taylor-Galerkin method is also tested using $\Delta t=2.0\times 10^{-4}s$ as a stable time step,  and a corresponding result is shown in \ref{Displacement of leaflet tip as a function of time}(b). This shows a similar magnitude (1.24) and frequency (2.86). These results are all within the range of values in \cite[Table 4]{Kadapa_2016}. Note that since the initial condition before oscillation for these simulations is an unstable equilibrium, the first perturbation from this regime is due to numerical disturbances. Consequently, the initial transient regimes observed for the two methods (Least-squares and Taylor-Galerkin methods) are quite different. It is possible that an explicit method causes these numerical perturbations more easily, therefore makes the leaflet start to oscillate at an earlier stage than when using Least-squares approach.

\subsection{Solid disc in a cavity flow}
This numerical example is used to compare our method with the IFEM, which is described in \cite{Wang_2009,Zhao_2008}. In order to compare in detail, we also implement the IFEM, but we implemented it on an adaptive mesh with hanging nodes, and we use the isoparametric FEM interpolation function rather than the discretized delta function or RKPM function of \cite{zhang2004immersed,Zhang_2007}. 

\begin{figure}[h!]
	\begin{minipage}[h!]{0.5\linewidth}
		\centering  
		\includegraphics[width=1.8in,angle=0]{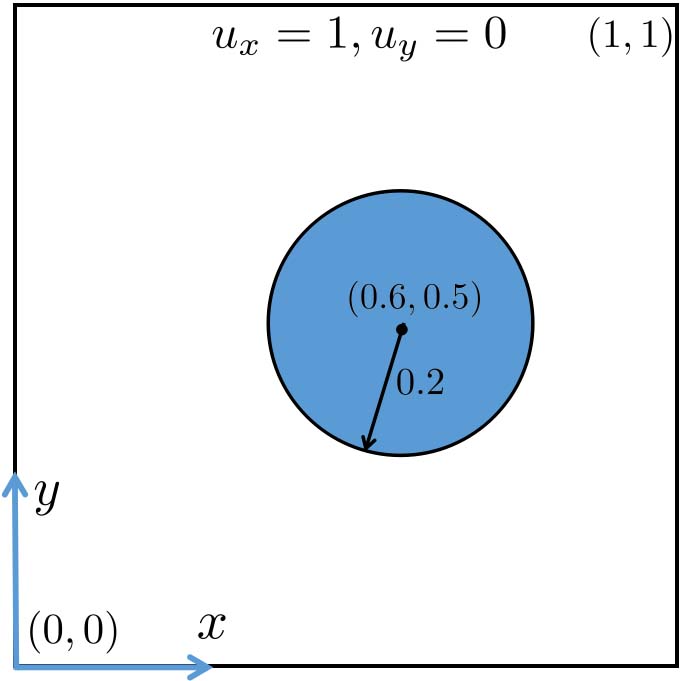}	
		\captionsetup{justification=centering}
		\caption {\scriptsize Computational domain for \\ cavity flow, taken from \cite{Zhao_2008}.} 
		\label{Computational domain for cavity flow}
	\end{minipage}
	\begin{minipage}[h!]{0.5\linewidth}
		\centering  
		\includegraphics[width=1.8in,angle=0]{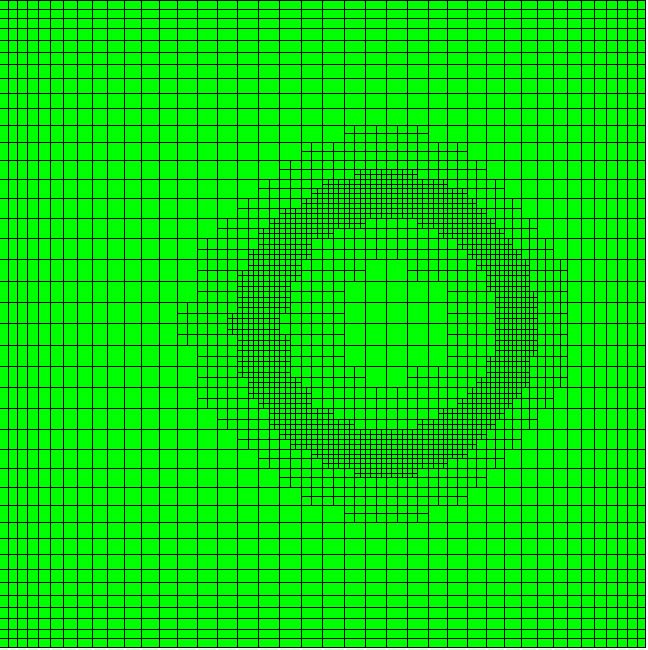}
		\captionsetup{justification=centering}	
		\caption {\scriptsize Adaptive mesh for \\ cavity flow.} 
		\label{Adaptive mesh for cavity flow}	
	\end{minipage}     			
\end{figure}

The fluid and solid properties are chosen to be the same as in \cite{Zhao_2008}: $\rho^f=\rho^s=1.0$, $\mu^f=0.01$ and $\mu^s=0.1$. The horizontal velocity on the top boundary of the cavity is prescribed as 1 and the vertical velocity is fixed to be 0 as shown in Figure \ref{Computational domain for cavity flow}. The velocities on the other three boundaries are all fixed to be 0, and pressure at the bottom-left point is fixed to be 0 as a reference point.

\begin{figure}[h!]
	\begin{minipage}[h!]{1\linewidth}
		\centering  
		\includegraphics[width=4.5in,angle=0]{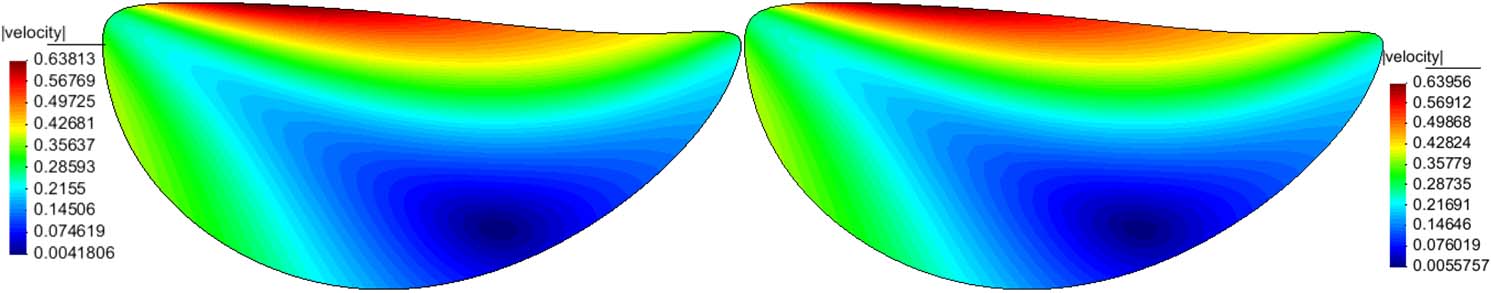}	
		\caption*{\scriptsize(a) $t=4.5s$.}
	\end{minipage}
	\begin{minipage}[h!]{1\linewidth}
		\centering  
		\includegraphics[width=4.5in,angle=0]{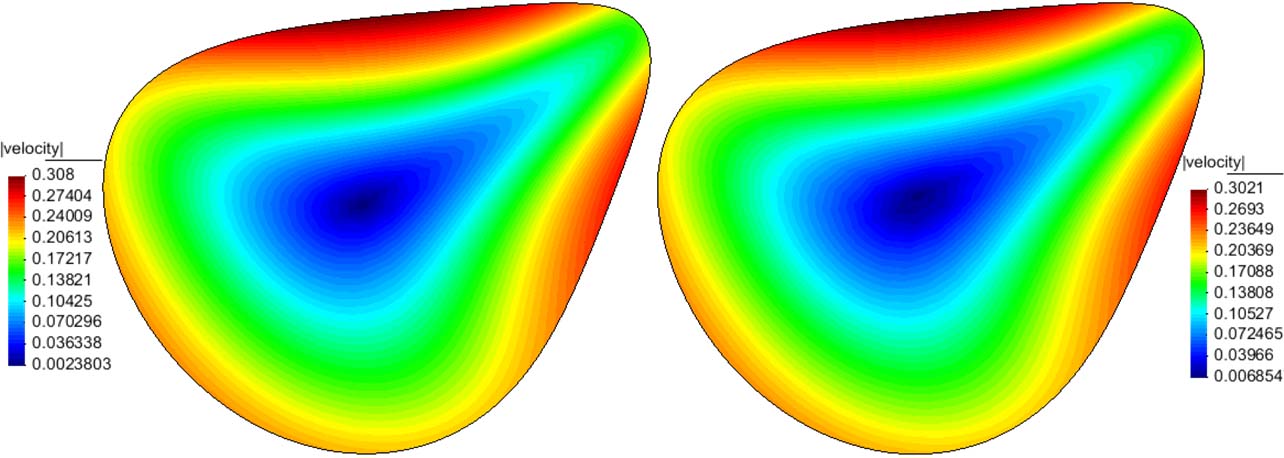}
		\caption*{\scriptsize(b) $t=25.5s$.}		
	\end{minipage}     		
	\captionsetup{justification=centering}
	\caption {\scriptsize Velocity norm for a soft solid $\left(\mu^s=0.1\right)$ in a driven cavity flow \\
		using our method (left) and IFEM (right).} 
	\label{ Velocity norm for a soft solid1}	
\end{figure}

In order to compare our method with IFEM we use the same meshes for fluid and solid: the solid mesh has 2381 nodes and the fluid mesh locally has a similar number of nodes (adaptive, see Figure \ref{Adaptive mesh for cavity flow}). First the Least-squares method is used to solve the convection step, and the time step is $\Delta t=1.0\times10^{-3}$. Figure \ref{ Velocity norm for a soft solid1} shows the configuration of the deformed disc at different stages, from which we do not observe significant differences of the velocity norm even for a long run as shown in Figure \ref{ Velocity norm for a soft solid1} (b). Then Taylor-Galerkin method is tested, and we achieve almost the same accuracy by using the same time step (not shown in the Figure).

We also test different densities, and the cases of $\mu^s=1.0$ and $\mu^s=100$. For our proposed method we can use $\mu^s=100$ or larger in order to make the solid behave like a rigid body without changing time step (again, not shown here due to lack of space). This is not possible for the IFEM for which the simulation always breaks down for $\mu^s=100$, however small the time step, due to the huge FSI force on the right-hand side of the IFEM system. 

\subsection{Falling disc in a channel with gravity}
\begin{figure}[h!]
	\centering  
	\includegraphics[width=4.5in,angle=0]{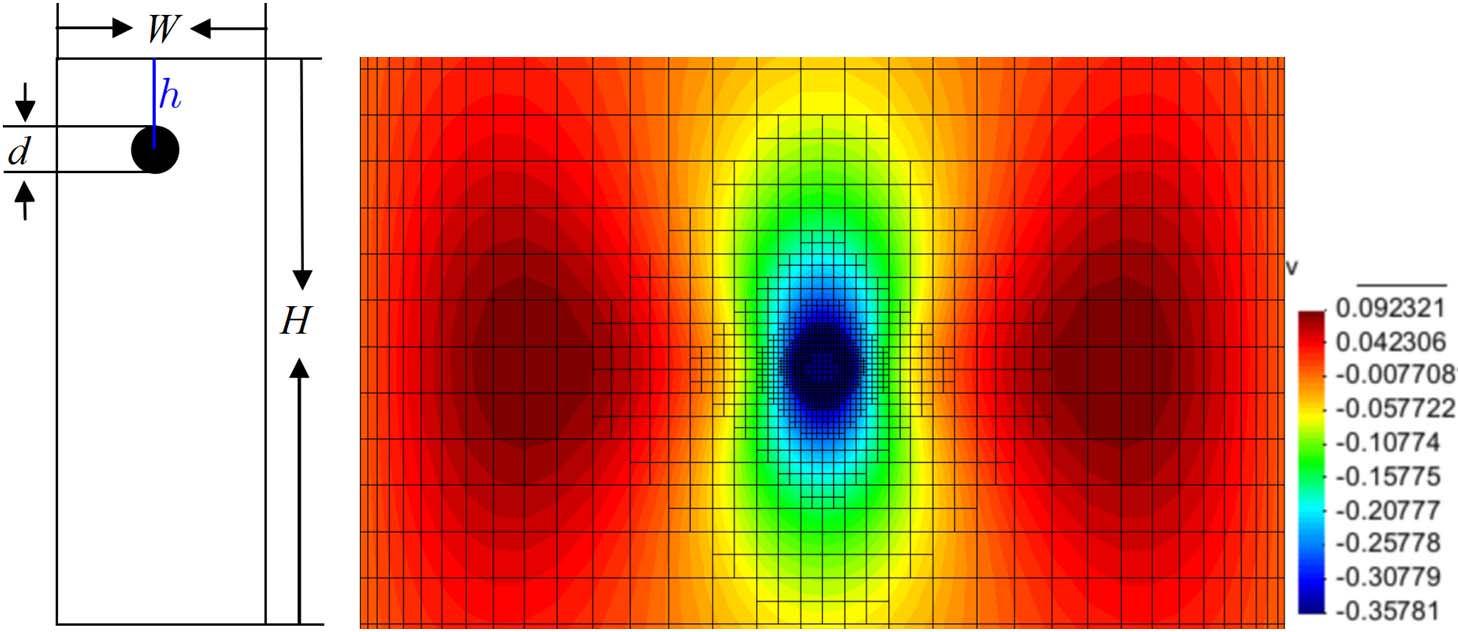}
	\caption* {\scriptsize (a) Computational domain \qquad
		 (b) Contour of vertical velocity at $t=1s$ (fine mesh)} 
	\caption {\scriptsize Falling disc in a channel with gravity.} 
	\label{Falling disc in a channel with gravity}
\end{figure}

The final test that we present in this paper is that of a falling disc in a channel, as cited by \cite{Zhang_2007,Hesch_2014} for example, in order to further validate against the IFEM and a monolithic method respectively. The computational domain and parameters are illustrated in Figure \ref{Falling disc in a channel with gravity} (a) and Table \ref{Fluid and material properties of a falling disc} respectively. The fluid velocity is fixed to be 0 on all boundaries except the top one.
\begin{table}[h!]
	\centering
	\begin{tabular}{|c|c|}
		\hline
		Fluid & Disc \\
		\hline
		$W=2.0$ $cm$ & $d=0.125$ $cm$ \\
		$H=4.0$ $cm$ & $h=0.5$ $cm$ \\
		$\rho^f=1.0$ $\left.g\right/{cm^3}$ & $\rho^s=1.2$ $\left.g\right/{cm^3}$ \\
		$\mu^f=1.0$ $\left.dyne\cdot s\right/{cm^2}$ &  $\mu^s=10^8$ $\left.dyne\right/{cm^2}$ \\
		$g=980$ $\left.cm\right/s^2$ & $g=980$ $\left.cm\right/s^2$ \\		
		\hline
	\end{tabular}
	\caption{Fluid and material properties of a falling disc.}
	\label{Fluid and material properties of a falling disc}
\end{table}

There is also an empirical solution of a rigid ball falling in a viscous fluid \cite{Hesch_2014}, for which the terminal velocity, $u_t$, under gravity is given by
\begin{equation}\label{empirical formula}
u_t=\frac{\left(\rho^s-\rho^f\right)gr^2}{4\mu^f}\left(ln\left(\frac{L}{r}\right)-0.9157+1.7244\left(\frac{r}{L}\right)^2-1.7302\left(\frac{r}{L}\right)^4\right),
\end{equation}
where $\rho^s$ and $\rho^f$ are the density of solid and fluid respectively, $\mu^f$ is viscosity of the fluid, $g=980$ $\left.cm\right/s^2$ is acceleration due to gravity, $\left.L=W\right/2$ and $r$ is the radius of the falling ball. We choose $\mu^s=10^8$ $\left.dyne\right/cm^2$ to simulate a rigid body here, and $\mu^s=10^{12}$ $\left.dyne\right/cm^2$ is also applied, which gives virtually identical results.

\begin{figure}[h!]
	\centering  
	\includegraphics[width=4in,angle=0]{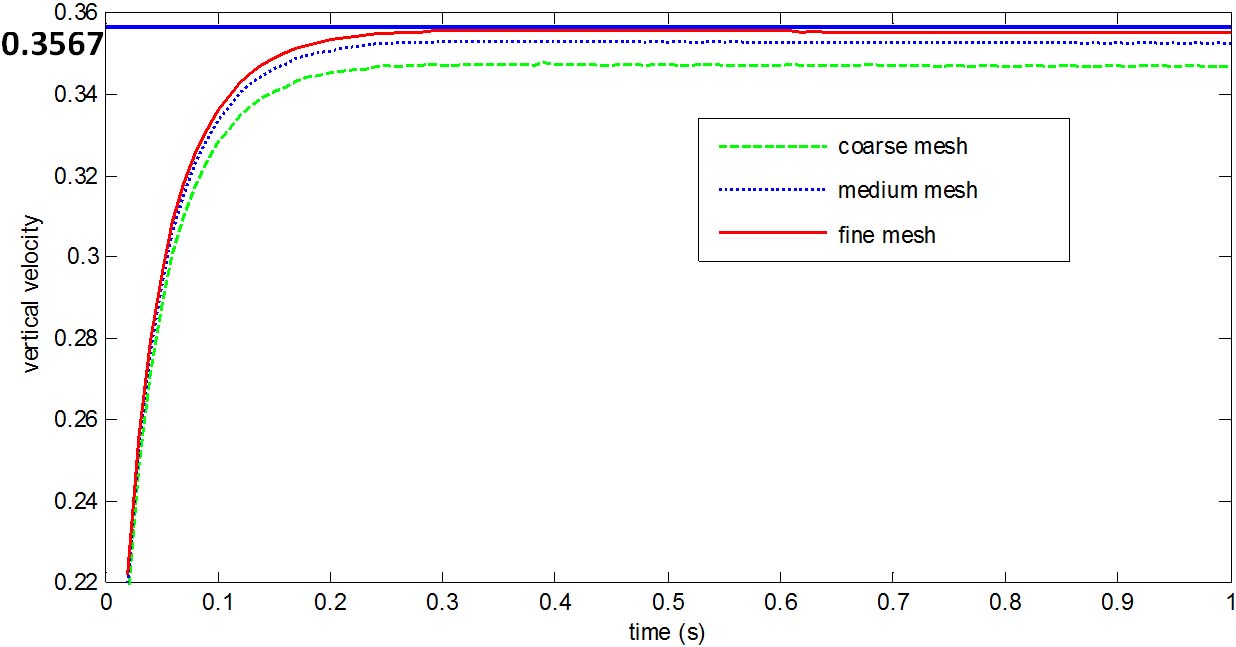}
	\captionsetup{justification=centering}
	\caption {\scriptsize Evolution of velocity at the center of a falling disc. \\
		(The blue solid line represents the empirical solution from formula (\ref{empirical formula}),\\
		for interpretation of the reference to color, the reader is referred to the web version.)} 
	\label{Evolution of velocity at center a falling disc}						
\end{figure}

Three different meshes are used: the disc boundary is represented with 28 nodes (coarse), 48 nodes (medium), or 80 nodes (fine). The fluid mesh near the solid boundary has the same mesh size as that of the disc, and a stable time step $t=0.005s$ is used for all three cases. A local snapshot of the vertical velocity with the adaptive mesh is shown in Figure \ref{Falling disc in a channel with gravity} (b). From the fluid velocity pattern around the disc we can observe that the disc behaves like a rigid body as expected. In addition, the evolution of the velocity of the mid-point of the disc is shown in Figure \ref{Evolution of velocity at center a falling disc}, from which it can be seen that the numerical solution converges from below to the empirical solution.

\section{Discussion}\label{Discussion}
In this section, some further remarks and notes concerning the proposed method are discussed.

\subsection{Treatment of the convection equation}
Both the Least-squares method and Taylor-Galerkin method add artificial diffusive terms in their fomulations to stablize the numerical scheme. Like all such stabilization approaches this necessarily has an influence on the accuracy, especially for large Reynolds numbers. In such cases a balance is required between minimizing the artificial dissipation and maintaining a stable time step size that is acceptable. In our applications, the Reynolds number is around $100\sim 500$, except for two extreme test cases in section \ref{Oscillation of a flexible leaflet oriented across the flow direction} whose Reynolds numbers are 1000 and 5000 respectively (Figure \ref{permutation of parameters} (b)). Even then, in these cases a minimal amount of diffusion is observed provided we use a small time step $(5\times10^{-4})$. Alternatively, an upwind scheme or a discontinuous Galerkin method could be a better choice. However, we have not yet implemented such methods on the adaptive mesh with hanging nodes.

\subsection{The Lagrangian update of the solid}
Updating the solid based upon its velocity could lead to distorted elements, either in its interior or at its boundary. Should this occur there are advanced mesh update techniques to improve the quality of solid mesh \cite{Bazilevs_2013} or discrete remeshing may be used \cite{peterson1999solution}. However all of the tests undertaken in this article have been performed based upon published benchmarks using incompressible solids and a small time step, and we have not encountered the problem of significantly distorted elements. In other applications our simple Lagrangian approach may not be adequate and so ALE techniques, possibly including mesh quality improvement, may also be required.

\subsection{Contact between solids and boundaries}
In many applications moving solids may run into boundaries (either external or of other moving bodies). In this article, we have only considered standard benchmark problems for which contact does not arise. Hence, through the use of a small time step and an adaptive algorithm to refine the mesh when the solids are near each other or near the boundaries, we have not needed to implement a contact test or a contact model. In the future, we do intend to consider adding a contact model in order to further generalize our method.

\subsection{Conditioning of the linear system}	
If $\rho^s\ge\rho^f$, and negleting $\tau_{ij}^f$, the discretized linear equation system is guaranteed to be well-conditioned. However, this restriction is too stringent to be a necessary condition. For example, we have implemented and tested a number of cases for which $\rho^s<\rho^f$ and the solid rises in a stable manner due to buoyancy.

\subsection{Approximation for pressure}
It is well known that the pressure jumps across the interface between the fluid and solid, and that a high resolution is therefore needed near the interface in order to capture this jump. In this article, we use an adaptive mesh refinement near the interface to reduce the error caused by our continuous approximation (${\bf P}_2{\rm P}_1$ element) for this discontinuous pressure. An alternative or additional choice is to  use ${\bf P}_2\left({\rm P}_1+{\rm C}\right)$ elements (the shape function of pressue is enriched by a constant) in order to capture an element-based jump of pressure. We intend to test this element in the future.

\section{Conclusion}\label{Conclusion and future works}
In this article we introduce a one-field FD method for fluid-structure interaction, which can be applied to a wide range of problems, from small deformation to very large deformation and from very soft solids through to very rigid solids. Several numerical examples, which are widely used in the literature of IFEM and FD methods with DLM (DLM/FD), are implemented to validate the proposed method. 

The one-field FD method combines features from both the IFEM and DLM/FD. Nevertheless, it differs from each of them in the following aspects. Firstly, our one-field FD method solves the solid and fluid equations together while the classical IFEM does not solve the solid equations. Although the implicit form of IFEM can iteratively solve the solid equations, this is different from our one-field FD method which couples the fluid and solid equations monolithically via a direct matrix addition as shown in formulas (\ref{matirx_A}) and (\ref{forcevector_b}). Secondly, while both our one-field FD method and DLM/FD solve solid equations, the former solves for just one velocity field in the solid domain using FEM interpolation, while the latter solves one velocity field and one displacement field in the solid domain using Lagrange multipliers. In summary therefore we believe that the one-field FD method has the potential to offer the robustness and range of operation of DLM/FD, but at a computational cost that is much closer to that of the IFEM approaches. Expressed another way, we contend that our approach has all of the advantages of IFEM techniques but the additional robustness usually associated with more complex monolithic solvers.

\appendix
\section{Expressions of ${\rm M}$, ${\rm M}^s$, ${\rm K}$, ${\rm K}^s$, ${\rm B}$, ${\rm f}$ and ${\rm f}^s$} \label{expressions_of_maatrices_and_vectors}
In this appendix, the specific expressions for the mass matrices ${\bf M}$ and ${\bf M}^s$, stiffness matrices ${\bf K}$ and ${\bf K}^s$, matrix ${\bf B}$ and the force vectors ${\bf f}$ and ${\bf f}^s$ in equations (\ref{lastlinerequation}), (\ref{matirx_A}) and (\ref{forcevector_b}) are presented.
\begin{enumerate}[(1)]
\item ${\bf M}$: $\left(k,m=1,2,\cdots N^u\right)$
	\begin{equation*}
	{\bf M}=\rho^f
	\begin{bmatrix}
	{\bf M}_{11} & {} \\
	{} & {\bf M}_{22} \\
	\end{bmatrix}
	, 	
	\left({\bf M}_{11}\right)_{km}=\left({\bf M}_{22}\right)_{km}
	=\left(\varphi_k,\varphi_m\right)_{\Omega^h}.
	\end{equation*}

\item  ${\bf M}^s$: $\left(k,m=1,2,\cdots N^s\right)$
\begin{equation*}
{\bf M}^s=\left(\rho^s-\rho^f\right)
\begin{bmatrix}
{\bf M}_{11}^s & {} \\
{} & {\bf M}_{22}^s \\
\end{bmatrix}
,
\left({\bf M}_{11}^s\right)_{km}
=\left({\bf M}_{22}^s\right)_{km}
=\left(\varphi_k^s,\varphi_m^s\right)_{\Omega^{sh}}.
\end{equation*}

\item  ${\bf K}$: $(k,m=1,2,\cdots N^u$)
\begin{equation*}
{\bf K}=\mu^f
\begin{bmatrix}
{{\bf K}_{11}} & {{\bf K}_{12}} \\
{{\bf K}_{21}} & {{\bf K}_{22}} \\
\end{bmatrix}
,
\end{equation*}
where
$$
\left({\bf K}_{11}\right)_{km}=2\left(\frac{\partial \varphi_k}{\partial x_1}, \frac{\partial \varphi_m}{\partial x_1}\right)_{\Omega^h}+\left(\frac{\partial \varphi_k}{\partial x_2}, \frac{\partial \varphi_m}{\partial x_2}\right)_{\Omega^h},
$$
$$
\left({\bf K}_{22}\right)_{km}=2\left(\frac{\partial \varphi_k}{\partial x_2}, \frac{\partial \varphi_m}{\partial x_2}\right)_{\Omega^h}+\left(\frac{\partial \varphi_k}{\partial x_1}, \frac{\partial \varphi_m}{\partial x_1}\right)_{\Omega^h},
$$
$$
\left({\bf K}_{12}\right)_{km}=\left(\frac{\partial \varphi_k}{\partial x_1}, \frac{\partial \varphi_m}{\partial x_2}\right)_{\Omega^h},
\left({\bf K}_{21}\right)_{km}=\left({\bf K}_{12}\right)_{mk}=\left(\frac{\partial \varphi_k}{\partial x_2}, \frac{\partial \varphi_m}{\partial x_1}\right)_{\Omega^h}.
$$

\item  ${\bf K}^s$: $\left(b,m=1,2,\cdots N^s\right)$
\begin{equation}
{\bf K}^s=
\begin{bmatrix}
{{\bf K}_{11}^s} & {{\bf K}_{12}^s} \\
{\bf K}_{21}^s & {{\bf K}_{22}^s} \\
\end{bmatrix},
\end{equation}
where 
\begin{equation*}
\begin{split}
&\left({\bf K}_{11}^s\right)_{bm}=\mu^s \Delta t2\left(\frac{\partial \varphi_b^s}{\partial x_1}, \frac{\partial \varphi_m^s}{\partial x_1}\right)_{\Omega^{sh}}+\mu^s \Delta t\left(\frac{\partial \varphi_b^s}{\partial x_2},
\frac{\partial \varphi_m^s}{\partial x_2}\right)_{\Omega^{sh}} \\
& +2\mu^s \Delta t^2\left(\frac{\partial \varphi_b^s}{\partial x_k}\frac{\partial u_1^n}{\partial x_k}, \frac{\partial \varphi_m^s}{\partial x_1}\right)_{\Omega^{sh}}+\mu^s \Delta t^2\left(\frac{\partial \varphi_b^s}{\partial x_k}\frac{\partial u_2^n}{\partial x_k},
\frac{\partial \varphi_m^s}{\partial x_2}\right)_{\Omega^{sh}} \\
&+ 2\Delta t^2\left(\frac{\partial \varphi_b^s}{\partial x_k}\left(\tau_{kl}^s\right)^n\frac{\partial u_1^n}{\partial x_l}, \frac{\partial \varphi_m^s}{\partial x_1}\right)_{\Omega^{sh}}+\Delta t^2\left(\frac{\partial \varphi_b^s}{\partial x_k}\left(\tau_{kl}^s\right)^n\frac{\partial u_2^n}{\partial x_l},
\frac{\partial \varphi_m^s}{\partial x_2}\right)_{\Omega^{sh}} \\
& +2\Delta t\left(\frac{\partial \varphi_b^s}{\partial x_k}\left(\tau_{k1}^s\right)^n, \frac{\partial \varphi_m^s}{\partial x_1}\right)_{\Omega^{sh}}+\Delta t\left(\frac{\partial \varphi_b^s}{\partial x_k}\left(\tau_{k2}^s\right)^n,
\frac{\partial \varphi_m^s}{\partial x_2}\right)_{\Omega^{sh}}.
\end{split}
\end{equation*}
${\bf K}_{22}^s$ can be expressed by changing the subscript $1$ to $2$ and $2$ to $1$ in the formula of ${\bf K}_{11}^s$.
\begin{equation*}
\begin{split}
&\left({\bf K}_{12}^s\right)_{bm}=\mu^s \Delta t\left(\frac{\partial \varphi_b^s}{\partial x_1},
\frac{\partial \varphi_m^s}{\partial x_2}\right)_{\Omega^{sh}}
+\mu^s \Delta t^2\left(\frac{\partial u_1^n}{\partial x_k}\frac{\partial \varphi_b^s}{\partial x_k},
\frac{\partial \varphi_m^s}{\partial x_2}\right)_{\Omega^{sh}}\\
&+\Delta t^2\left(\frac{\partial u_1^n}{\partial x_k}\left(\tau_{kl}^s\right)^n\frac{\partial \varphi_b^s}{\partial x_l},
\frac{\partial \varphi_m^s}{\partial x_2}\right)_{\Omega^{sh}}
+\Delta t\left(\left(\tau_{1k}^s\right)^n\frac{\partial \varphi_b^s}{\partial x_k},
\frac{\partial \varphi_m^s}{\partial x_2}\right)_{\Omega^{sh}},
\end{split}
\end{equation*}
and $\left({\bf K}_{21}^s\right)_{bm}=\left({\bf K}_{12}^s\right)_{mb}$.

\item  ${\bf B}$: $\left(k=1,2,\cdots N^p\right. $ and $\left. m=1,2,\cdots N^u\right)$
\begin{equation*}
{\bf B}=
\begin{bmatrix}
{\bf B}_1 \\
{\bf B}_2 \\
\end{bmatrix}
,
\left({\bf B}_i\right)_{mk}=-\left(\phi_k, \frac{\partial \varphi_m}{\partial x_i}\right)_{\Omega^h}, (i=1,2).
\end{equation*}

\item ${\bf f}$: $\left(m=1,2,\cdots N^u\right)$
\begin{equation*}
{\bf f}=
\begin{pmatrix}
{\bf f}_1 \\
{\bf f}_2 \\
\end{pmatrix}
,
\left({\bf f}_i\right)_m=\rho^f\left(g_i,\varphi_m\right)_{\Omega^h}+\left(\bar{h}_i,\varphi_m\right)_{\Gamma^{Nh}}, 
(i=1,2)
\end{equation*}

\item ${\bf f}^s$: $\left(m=1,2,\cdots N^s\right)$
\begin{equation*}
\begin{split}
& {\bf f}^s=
\begin{pmatrix}
{\bf f}_1^s \\
{\bf f}_2^s \\
\end{pmatrix}
,
\left({\bf f}_i^s\right)_m=\left(\rho^s-\rho^f\right)\left(g_i,\varphi_m^s\right)_{\Omega^{sh}} \\
&+\left(\mu^s\Delta t^2\frac{\partial u_i^n}{\partial x_k}\frac{\partial u_j^n}{\partial x_k} 
+\Delta t^2\frac{\partial u_i^n}{\partial x_k}\left(\tau_{kl}^s\right)^n\frac{\partial u_j^n}{\partial x_l}-\left(\tau_{ij}^s\right)^n, \frac{\partial \varphi_m^s}{\partial x_j}
\right)_{\Omega^{sh}}, (i=1,2).
\end{split}
\end{equation*}

\end{enumerate}


\end{document}